\documentclass[12pt, final, onecolumn ]{IEEEtran}

\usepackage{chngcntr}

\usepackage{graphicx}
\usepackage{color}

 \usepackage{amsmath}
\usepackage{setspace}
\usepackage{enumerate}
\usepackage{verbatim}
\usepackage{amssymb}
\usepackage{amsbsy}
\usepackage{theorem}
\usepackage{url}

\usepackage[dvips]{epsfig}

\def\urltilde{\kern -.15em\lower .7ex\hbox{\~{}}\kern .04em}
\def\urldot{\kern -.10em.\kern -.10em}

\def\urlhttp{http\kern -.10em\lower -.1ex\hbox{:}\kern -.12em\lower 0ex\hbox{/}\kern -.18em\lower 0ex\hbox{/}}

\theoremstyle{plain}
\newtheorem{Theorem}{Theorem}

\newtheorem{Proposition}{Proposition}

\newtheorem{Problem}{Problem}
\newtheorem{Fact}{Fact}
{\theorembodyfont{\rmfamily} \newtheorem{Example}{Example}}
\newtheorem{Remark}{Remark}

\newcommand {\R}{\mathbb R}
\newcommand {\U}{\mathbb U}

\newcommand{\be}{\begin{equation}}
\newcommand{\ee}{\end{equation}}
\newcommand{\Int}{\operatorname{{\mathrm int}}}
\newcommand{\ad}{\operatorname{{\mathrm ad}}}

{\theorembodyfont{\rmfamily} \newtheorem{Definition}{Definition}}

 \newcommand{\profile}{augmented profile}

\newcommand{\LMD}{\lambda_0,\dots,\lambda_n}

\doublespace
\begin{document}
\title{Controllability analysis and control synthesis for the ribosome flow model\thanks{This research is partially supported by  research grants
 from  the Israeli Science Foundation, the Israeli Ministry of Science, Technology \& Space, the Edmond J. Safra Center for Bioinformatics at Tel Aviv University and the US-Israel Binational Science Foundation. The work of EDS is also supported in part by grants AFOSR FA9550-14-1-0060 and ONR 5710003367.\newline
An abridged version of this paper has been  presented at
the \emph{55th IEEE Conference on Decision and Control}~\cite{cdc_rfm_cont2016}.
}}
\author{ Yoram Zarai,   Michael Margaliot,   Eduardo D. Sontag and Tamir Tuller* \IEEEcompsocitemizethanks{
\IEEEcompsocthanksitem
 Y. Zarai is with the School of Electrical Engineering, Tel-Aviv
University, Tel-Aviv 69978, Israel.
E-mail: yoramzar@mail.tau.ac.il
\IEEEcompsocthanksitem
M. Margaliot is with the School of Electrical Engineering and the Sagol School of Neuroscience, Tel-Aviv
University, Tel-Aviv 69978, Israel.
E-mail: michaelm@eng.tau.ac.il
\IEEEcompsocthanksitem
E.  D. Sontag is with the Dept. of Mathematics and Cancer Center of New Jersey, Rutgers
  University, Piscataway, NJ 08854, USA. E-mail:  eduardo.sontag@gmail.com
\IEEEcompsocthanksitem
  T. Tuller is with the Dept. of Biomedical Engineering and the Sagol School of Neuroscience, Tel-Aviv
University, Tel-Aviv 69978, Israel.
E-mail: tamirtul@gmail.com
\IEEEcompsocthanksitem
*Corresponding authors: TT and EDS.
 }}

\maketitle

\begin{abstract}
The ribosomal density
 along different parts of the coding regions of the mRNA molecule affects various fundamental intracellular phenomena including: protein production rates, global ribosome allocation and organismal fitness, ribosomal drop off, co-translational protein folding, mRNA degradation, and more. Thus, regulating translation in order to obtain
   a \emph{desired}  ribosomal  profile along the mRNA molecule is an
 important biological problem.

We study this problem by using  a dynamical   model for mRNA translation, called the ribosome flow model~(RFM).
 In the RFM, the mRNA molecule is modeled as an ordered chain of~$n$ sites.
 The RFM includes~$n$ state-variables describing  the ribosomal density profile along the mRNA molecule,
 and  the transition rates from each site to the next are controlled
by~$n+1$    positive constants.
To study the problem of controlling the density profile, we consider  some or all of the
  transition rates as  time-varying controls.

We  consider  the following problem:
given an initial  and a  desired ribosomal density profile in the RFM,  determine
the time-varying values of the transition rates that steer the system to the desired
 density profile, if they exist.
More specifically, we  consider two control problems. In the first, all transition rates can be  regulated separately,
and the goal is to steer the ribosomal density profile and the protein production rate from a given initial value
to a desired value.
In the second problem, one or more transition rates are jointly regulated  by a single scalar control,
 and the goal is to steer the production rate to
a desired value within a certain  set of feasible values. In the first case,
  we show that the system is controllable, i.e.
the control is powerful enough to steer the system to any desired value in finite time, and   provide
simple closed-form  expressions for \emph{constant} positive control functions (or transition rates) that asymptotically
steer the system to the  desired value. In the second case,  we show that the system is controllable,
and provide a simple algorithm for determining
the \emph{constant} positive  control value that asymptotically
steers the system to the  desired value.
 We discuss some of the biological implications of these results.
 \end{abstract}

\section{Introduction}
The process in which the genetic information coded  in the DNA is transformed into functional proteins is called \emph{gene expression}. It consists of two major steps: \emph{transcription} of the DNA code into messenger RNA (mRNA) by RNA polymerase, and \emph{translation} of the mRNA into proteins. During the translation step, complex  macro-molecules called ribosomes unidirectionally traverse the mRNA, decoding it codon by codon into a corresponding chain of amino-acids that
 is folded co-translationally and post-translationally to become a functional protein.
 The rate in which proteins are produced during the translation step is called the \emph{protein translation rate} or \emph{protein production rate}.

Translation takes place
 in all living organisms and all tissues under almost all conditions. Thus, developing a better understanding of how translation is regulated  has important implications to many scientific disciplines, including medicine, evolutionary biology, and synthetic biology. Developing and analyzing computational models of translation may provide important insights on this biological process. Such models can also aid in integrating and analyzing the rapidly increasing experimental findings related to translation~(see, e.g., \cite{Dana2011,TullerGB2011,Tuller2007,Chu2012,Shah2013,Deneke2013,Racle2013}).

Controlling the expression of heterologous genes in a host organism in order to synthesize new proteins, or to improve certain aspects of the host fitness, is an essential challenge in biotechnology and synthetic biology~\cite{Romanos1992,Moks1987,Binnie1997,Tuller2015,Arava2003}.
Computational models of translation are particularly important in this context, as they allow simulating and analyzing the effect of various manipulations of the gene expression machinery and/or the genetic material,
and can thus save considerable time and effort by  guiding biologists towards promising  experimental directions.

The ribosome flow along the mRNA is regulated   by various translation factors
 (e.g.,
   initiation and elongation factors, tRNA and Aminoacyl tRNA synthetase concentrations, and amino-acid concentrations) in order to achieve both a suitable  ribosomal density profile  along the mRNA, and a desired   protein production rate.
Indeed,
it is known that the ribosomal density profile and the induced ribosome speed profile along the mRNA molecule
can affect various fundamental intracellular phenomena. For example,
 it is known that  the folding of   translated proteins may take place    co-translationally,
  and inaccurate translation speed can contribute to protein mis-folding~\cite{Drummond2008,Kimchi-Sarfaty2013,Zhang2009}.
	The ribosome density profile also  affects the degradation of mRNA, ribosomal collisions, abortion and allocation, transcription, and more \cite{Drummond2008,Kimchi-Sarfaty2013,Kurland1992,Edri2014,Zhang2009,Tuller2015,Proshkin2010}.

Thus, a natural question is whether it is   possible, by controlling  the   transition  rates  along the mRNA,
   to steer the ribosome density   along the mRNA molecule  from any initial
profile to any desired profile in finite time, and if so, how.
 In the language of control theory, the question is whether  the system is~\emph{controllable}
(see, e.g.~\cite{sontag_book}), and if so, how to solve the control synthesis problem.
We note that
 controllability of networked systems is recently attracting considerable
interest (see e.g.~\cite{barabashi_nature}).
 Controllability of such networks depends on the interplay between two factors: (1)~the network's 
topology,
 and (2)~the dynamical rules describing the behavior at each network node.
When studying real-world networks, many of the parameter values in the network are not known explicitly.
The network is said to be \emph{structurally stable}
 if it will be controllable for almost every random selection of 
 parameter  values~\cite{lin1974structural,siljak_book,sontag_book,strong_struct_control}.
 An important problem in this context is to determine a minimal
set of ``driver nodes''
within the network such that controlling these nodes
 makes the entire network controllable or structurally controllable
(see e.g.~\cite{olshevsky_minimal_control,barabashi_nature}).

 Controllability of mRNA translation  is also important in   synthetic biology, e.g.
  in order to design    cis or trans intra-cellular elements that yield a desired
	ribosome density profile (or to determine if such a design is possible).
  Another related question arises in
  evolutionary systems biology, namely,
  determine if a certain translation-related phenotype   can be obtained by evolution.

The ribosome density profile is     also related to cancer evolution. Indeed, it
 is well-known that cancerous cells undergo evolution that modulates
 their translation regime. It has been  suggested that various   mutations that accumulate during tumorigenesis may affect both   translation initiation~\cite{nature_cancer_12,Loayza-Puch2013}
 and   elongation \cite{can_tuller_2009,Tomlinson2005}
 of genes related to cell proliferation, metabolism, and invasion.
 Specifically, the results reported in~\cite{nature_cancer_12} support the conjecture that cancerous mutations can significantly change the ribosome density profile on the mRNAs of dozens of genes.

The standard mathematical model for ribosome flow
is the
  \emph{totally asymmetric simple exclusion process}~(TASEP) \cite{Shaw2003,TASEP_tutorial_2011}.
 In this model,  particles   hop unidirectionally
 along an ordered lattice of~$L$ sites. Every site can be either free or occupied by a particle, and a particle can only
   hop to a free site. This  simple exclusion principle
   models particles that have ``volume'' and thus cannot overtake one  other.
   The hops are stochastic and the rate of hoping from site~$i$ to site~$i+1$  is denoted by~$\gamma_i$.
   A particle can hop to [from] the first [last] site of the lattice at a rate~$\alpha$   [$\beta$].
   The flow through the lattice converges to a steady-state value that depends on~$L$ and the parameters~$\alpha, \gamma_1,\dots,\gamma_{L-1},\beta$.
   In the context of translation, the lattice  models the mRNA molecule, the particles are ribosomes,
   and simple exclusion means that a ribosome  cannot overtake a ribosome in front of it.
    TASEP has become a
 fundamental model in non-equilibrium statistical mechanics, and has been
 applied to model numerous natural and artificial  processes~\cite{TASEP_book}.

The \emph{ribosome flow model}~(RFM)~\cite{reuveni} is a \emph{deterministic} model for mRNA translation that can be derived via  a dynamic
 mean-field approximation of TASEP~\cite[section 4.9.7]{TASEP_book}~\cite[p. R345]{solvers_guide}. In the RFM, mRNA molecules are coarse-grained into~$n$ consecutive sites of codons (or groups of codons). The state variable $x_i(t): \R_+ \to [0,1]$, $i=1,\dots,n$, describes the normalized ribosomal occupancy level (or density) of site~$i$ at time~$t$, where $x_i(t)=1$ [$x_i(t)=0$]
 indicates that site $i$ is completely full [empty] at time $t$.
Thus, the vector~$\begin{bmatrix}x_1(t)&\dots&x_n(t)\end{bmatrix}'$ describes the complete
\emph{ribosomal density profile} along the mRNA molecule at time~$t$.  A variable denoted~$R(t)$ describes
 the \emph{protein production rate} at time~$t$.
A non-negative parameter~$\lambda_i$, $i=0,\dots,n$, controls the transition rate from site~$i$ to site~$i+1$, where~$\lambda_0$ [$\lambda_n$] is
the initiation [exit] rate.

%

In order to better understand how translation is regulated,
 we
consider the RFM with some or all of the constant   transition  rates  replaced by   time-varying control functions that take
non-negative values for all time~$t$.
The idea here is that we can manipulate  these functions as desired.

We consider two control problems.
In the first, all the~$n+1$~$\lambda_i$s are replaced by control functions and the problem is to
manipulate  these functions such that both  the ribosomal density profile and the production rate are steered from a given initial value
to a  desired value. We use the term  ``{\profile}'' to indicate the combination of the ribosomal density profile and the production rate.

In the second control problem, we assume that all the rates belonging to some \emph{subset} of the rates
 are \emph{jointly} replaced by a single, scalar control~$u(t)$. We define  a set of ``relevant''
possible production rates and the 
  problem is to determine~$u(t)$ such that the production rate is steered to a desired value in this set.
Note that in the first problem
the~$(n+1)$-dimensional vector describing   the {\profile} is controlled using~$n+1$ control functions, and
  in the second problem  one variable is controlled using a scalar control.

We show that in both cases the resulting control system is \emph{controllable}, i.e.
the control is always ``powerful''
enough to steer the system  from any initial state to any desired state in some finite time~$T$.
	We also show that there always exists a control that  steers the system as desired, and is
   the time concatenation of two controls:
\be\label{eq:ucontca}
                u(t)=\begin{cases}
                        v ,& t \in [0,T-\varepsilon),\\
                        w(t),& t \in [T-\varepsilon, T],\end{cases}
\ee
with $\varepsilon>0$ and very small.
The constant control~$v$ is given in a \emph{simple  and explicit}  expression that depends only on  the desired final state.
 It guarantees
   that  this state becomes the unique attracting  steady-state ribosomal density and production rate of the RFM dynamics.
	For example, in the problem of controlling the density profile and the production rate to desired final values~$x^f$ and $R^f$ respectively (``$f$'' for final),
  the solution of the  controlled  RFM for \emph{any} initial condition~$x(0)$ and~$R(0)$ satisfies
\begin{align}\label{eq:limyhere}
\lim_{t\to\infty} x(t,v)&=x^f, \nonumber \\
\lim_{t\to\infty} R(t,v) &=R^f.
\end{align}
	This means that for all practical reasons, one may simply apply the constant control~$u(t)\equiv v$ for all~$t\geq0$.
Note that~\eqref{eq:limyhere}
 means  that the exact values of~$x(0)$ and~$R(0)$, i.e. the initial values of the density profile and production rate, are actually not needed.
This is important, as accurately
measuring~$x(0)$ and~$R(0)$  in practice may be   difficult.
The control~$w(t)$ in~\eqref{eq:ucontca} is needed only to guarantee that~$x(T)=x^f$ and $R(T)=R^f$ at the \emph{finite} time~$T$.
The existence of such a~$w(t)$ follows from Lie-algebraic
 accessibility arguments, but~$w(t)$ is not given explicitly.

Different aspects of translation regulation, usually under  natural conditions, have been studied before
 (see, for example,~\cite{Jackson2010}). There are also several
 studies on experimental and computational heuristics for mRNA translation engineering and optimization (see, for example, \cite{Salis2009,Sun2015}), and
 studies related to the way translation  regulation is encoded in the transcript (e.g. \cite{Zur2013,Plotkin2010}). However, to the best of our knowledge, this is the first study   on controllability and control synthesis
  in  a realistic dynamical model for translation. Also,
  previous studies on translation optimization only considered
	protein levels or production rate (e.g.~\cite{Salis2009}), but not the problem of controlling the
	entire profile of ribosome densities via changing the codon decoding rates, as is  done here.

The remainder of this paper  is organized as follows. The following section provides a brief overview of the RFM and its generalizations into
   a  control system. In order to make this paper accessible to a larger audience, Appendix~A provides
    a very brief review of     controllability, while demonstrating some of the concepts using the~RFM.
   Section~\ref{sec:main} presents our main results on  the controlled RFM. We also discuss the biological ramifications of our results.
   To streamline the presentation, all
 the proofs are placed in   Appendix~B.
We use standard notation. Vectors [matrices] are denoted by small [capital] letters. For a  vector~$x\in\R^n$,
$x_i$ is the~$i$th entry of~$x$, and~$x'$ is the transpose of~$x$. $\R^n_+$ [$\R^n_{++}$] is the set all $n$-tuples of nonnegative [strictly positive] real numbers.

\section{Ribosome Flow Model}

\begin{figure*}[t]
\centering
\includegraphics[height=3.2cm]{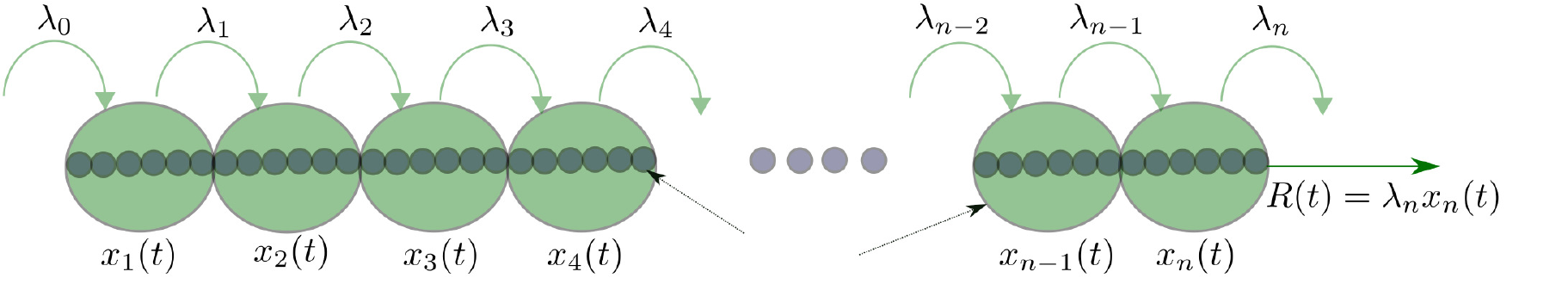}
\caption{The RFM models  a chain of $n$ sites of codons (or groups of codons).
The state variable~$x_i(t)\in[0,1]$ represents
 the normalized ribosome occupancy at site $i$ at time $t$.
 The elongation  rate  from  site~$i$ to site~$i+1$ is~$\lambda_i$,
with~$\lambda_0$ [$\lambda_n$] denoting  the initiation [exit] rate.
The production rate at time $t$ is $R(t) =\lambda_n x_n(t)$.}\label{fig:rfm}
\end{figure*}

In this section, we quickly review the RFM and   describe
its generalizations  into a control system.
 The dynamics of the RFM with $n$ sites is given by $n$ nonlinear first-order ordinary differential equations:
\begin{align}\label{eq:rfm}
                    \dot{x}_1&=\lambda_0 (1-x_1) -\lambda_1 x_1(1-x_2), \nonumber \\
                    \dot{x}_2&=\lambda_{1} x_{1} (1-x_{2}) -\lambda_{2} x_{2} (1-x_3) , \nonumber \\
                    \dot{x}_3&=\lambda_{2} x_{ 2} (1-x_{3}) -\lambda_{3} x_{3} (1-x_4) , \nonumber \\
                             &\vdots \nonumber \\
                    \dot{x}_{n-1}&=\lambda_{n-2} x_{n-2} (1-x_{n-1}) -\lambda_{n-1} x_{n-1} (1-x_n), \nonumber \\
                    \dot{x}_n&=\lambda_{n-1}x_{n-1} (1-x_n) -\lambda_n x_n.
\end{align}
If we define~$x_0(t):=1$ and $x_{n+1}(t):=0$
then~\eqref{eq:rfm} can be written more succinctly as
\be\label{eq:rfm_all}
\dot{x}_i=\lambda_{i-1}x_{i-1}(1-x_i)-\lambda_i x_i(1-x_{i+1}),\quad i=1,\dots,n.
\ee
 Recall that the state variable $x_i(t): \R_+ \to [0,1]$  describes the normalized ribosomal occupancy level (or density) at site $i$ at time~$t$, where $x_i(t)=1$ [$x_i(t)=0$]
 indicates that site $i$ is completely full [empty] at time~$t$.
Eq.~\eqref{eq:rfm_all}  can be explained as follows.
The flow of ribosomes from site~$i$ to site~$i+1$ is~$\lambda_{i} x_{i}(t)
(1 - x_{i+1}(t) )$. This flow is proportional to $x_i(t)$, i.e. it increases
with the occupancy level at site~$i$, and to $(1-x_{i+1}(t))$, i.e. it decreases as site~$i+1$ becomes fuller.  This corresponds to a ``soft'' version of the simple  exclusion principle in TASEP.
Note that the maximal possible  flow  from site~$i$ to site~$i+1$  is the transition rate~$\lambda_i$.
 Eq.~\eqref{eq:rfm_all} thus states that the
 time derivative
of  state-variable~$x_i$   is the  flow entering
site~$i$ from site~$i-1$,     minus the flow exiting site~$i$ to site~$i+1$.

The ribosome  exit rate from site $n$ at time $t$ is equal to the protein  production rate at
time~$t$, and is denoted by $R(t):=\lambda_n x_n(t)$ (see Fig.~\ref{fig:rfm}).
Note that~$x_i$ is dimensionless, and every rate~$\lambda_i$ has units of~$1/{\text{time}}$.

A system where each state variable describes the amount of ``material'' in some compartment, and the dynamics
describes the flow of material between  the compartments and also to/from  the surrounding
environment is called
a \emph{compartmental  system}~\cite{comp_surv_1993}. Compartmental  systems proved to be useful models
in  various biological domains   including physiology, pharmacokinetics, population dynamics, and epidemiology~\cite{comp_epi,comp_drug_sur,comp_mod_book}.
The  RFM is thus   a nonlinear  compartmental  model, with~$x_i$ denoting the normalized
amount of ``material'' in  compartment~$i$, and the flow follows  a ``soft'' simple exclusion  principle.
The controllability  of \emph{linear}  compartmental  systems
has been addressed in several papers~\cite{Johnson1976181,struc_cont_com_sys}.

Let~$x(t,a)$ denote the solution of~\eqref{eq:rfm}
at time~$t \ge 0$ for the initial
condition~$x(0)=a$. Since the  state-variables correspond to normalized occupancy levels,
  we always assume that~$a$ belongs to the  closed $n$-dimensional
  unit cube:
\[
           C^n:=\{x \in \R^n: x_i \in [0,1] , i=1,\dots,n\}.
\]
It has been  shown in~\cite{RFM_stability} that if~$a\in C^n$ then~$x(t,a) \in C^n$ for all~$t\geq0$, that is,~$C^n$ is an invariant set of the dynamics.
Let~$\Int(C^n)$ denote the interior of~$C^n$, and let~$\partial C^n$ denote the boundary of $C^n$.
Ref.~\cite{RFM_stability} has also shown that the RFM is a
\emph{tridiagonal cooperative dynamical system}~\cite{hlsmith},
and    that~\eqref{eq:rfm}
admits a \emph{unique} steady-state point~$e(\LMD) \in \Int(C^n)$ that is globally asymptotically stable, that is, $\lim_{t\to\infty} x(t,a)=e$  for all $a\in C^n$ (see also~\cite{RFM_entrain}). This means that the ribosome density profile always converges to a steady-state profile that depends on the rates, but not  on the initial condition. In particular, the production rate~$R(t)=\lambda_n x_n(t)$
 converges to a steady-state   value:
\be \label{eq:defr}
R:=\lambda_n  {e}_n.
\ee

At steady-state (i.e, for~$x=e$), the left-hand side of all the equations
in~\eqref{eq:rfm} is zero, so
\begin{align} \label{eq:ep}
                      \lambda_0 (1- {e}_1) & = \lambda_1 {e}_1(1- {e}_2)\nonumber \\&
                      = \lambda_2  {e}_2(1- {e}_3)   \nonumber \\ & \vdots \nonumber \\
                    &= \lambda_{n-1} {e}_{n-1} (1- {e}_n) \nonumber \\& =\lambda_n  {e}_n\nonumber\\&=R.
\end{align}
This yields
\begin{align}\label{eq:rall}
R=\lambda_i e_i(1-e_{i+1}), \quad i=0,\dots,n,
\end{align}
where $e_0:=1$ and $e_{n+1}:=0$.

\begin{Remark}\label{rem:Linvert}
One may view~\eqref{eq:ep}  as  a mapping from  the rates~$\begin{bmatrix} \lambda_0,\dots,\lambda_n\end{bmatrix}'$
to the steady-state density profile and production rate~$ \begin{bmatrix} e_1&\dots&e_n & R\end{bmatrix}'$.
For the purposes of  this paper, it is important to note that this mapping is \emph{invertible}. Indeed, Eq.~\eqref{eq:rall} implies that
 {given} a desired   density profile and production rate~$ \begin{bmatrix} e_1&\dots&e_n & R\end{bmatrix}'\in(0,1)^n\times \R_{++}$  one can immediately
 determine the transition  rates that yield this profile at steady-state, namely,
 \begin{align}\label{eq:inv}
\lambda_i&=\frac{R}{  e_i(1-e_{i+1})  } ,\quad i=0,\dots,n,
\end{align}
where~$e_0:=1$, and~$e_{n+1}:=0$.

Note that~\eqref{eq:inv} implies that~$\lambda_i$ increases with~$R$  and~$e_{i+1}$, and decreases  with~$e_i$.
This is intuitive, as  a larger~$\lambda_i$ implies a larger rate of ribosome
 flow from site~$i$ to site~$i+1$, as well as an increase in the steady-state
 production rate~\cite{RFM_max}.
 Thus, given a desired profile with  larger~$R$  and~$e_{i+1}$, and a smaller~$e_i$,
 the required transition  rates include a larger value for~$\lambda_i$.
\end{Remark}

\begin{figure}[t]
  \begin{center}
 \includegraphics[height=10cm]{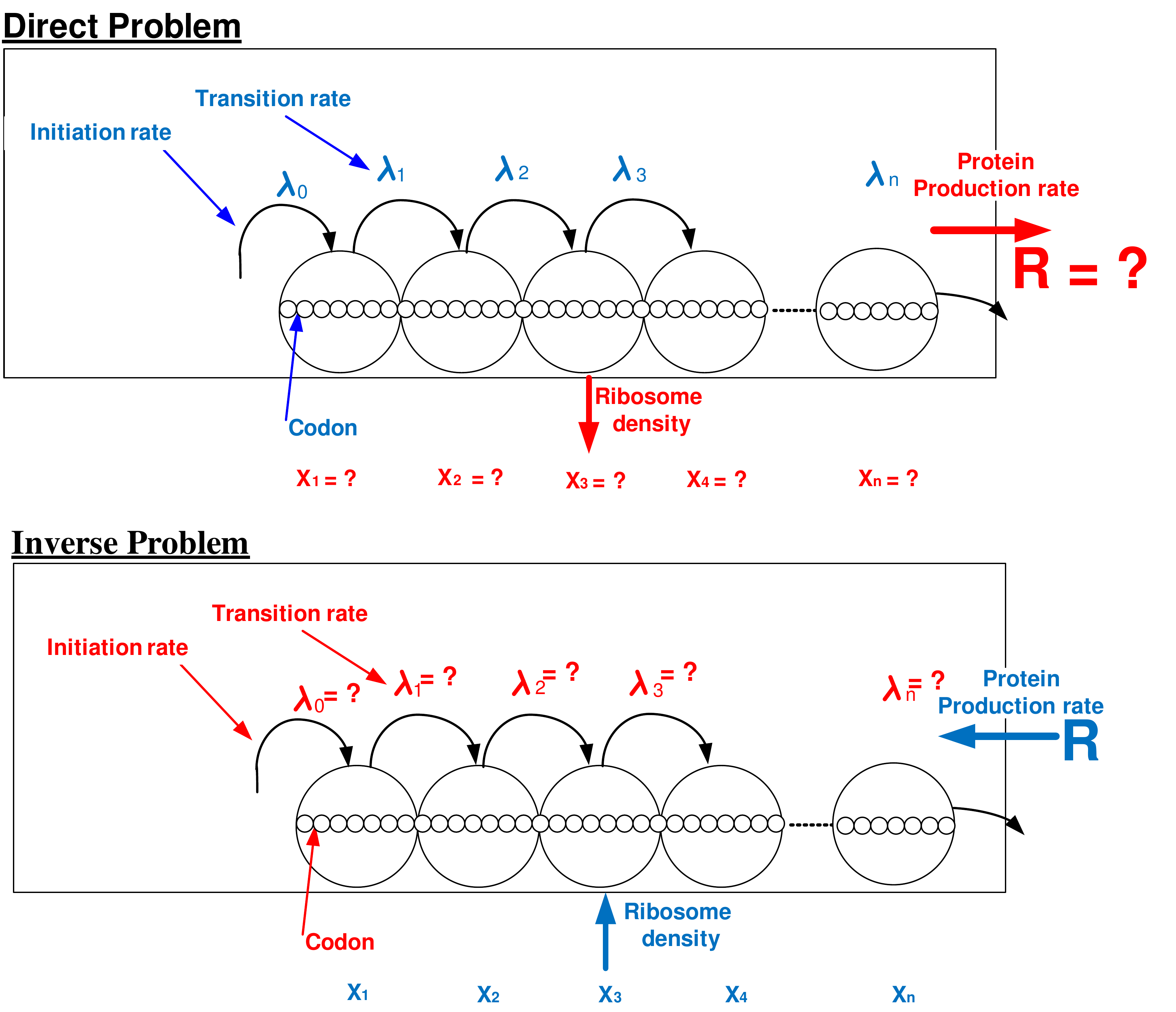}
  \caption{Upper part: Previous studies considered the direct problem:
	 given the RFM parameters, i.e. the set of transition rates $\lambda_i$s, analyze
	the dynamics of the RFM  ribosome densities~$x_{i}$s, and the production rate $R$.
  Lower part: here we consider the inverse problem: given a desired profile of ribosomal densities $x_{i}$, $i=1,\dots,n$, and a desired
	production
	rate~$R$,   find the   rates      that steer
	the dynamics  to this  profile.  } \label{fig:motivation}
  \end{center}
\end{figure}

From a biophysical point of view, this means that if there are
no  constraints on the transition rates then  we can engineer  any desired density profile together with a desired production rate.
More importantly, this provides an explicit expression for the needed rates. In addition to applications in
 functional genomics and molecular evolution, the observation in Remark~\ref{rem:Linvert} is also  related to problems in  synthetic biology  where
the goal is to re-engineer the mRNA molecule so as
to obtain a desired density profile and production rate (see Fig.~\ref{fig:motivation}).

For more on the analysis of the RFM using tools from systems and control theory, see~\cite{zarai_infi,RFM_max,RFM_sense,RFM_feedback,RFMR,RFM_r_max_density}.
The RFM models translation on a single isolated mRNA molecule. A network of~RFMs, interconnected through
 a common pool of ``free'' ribosomes has been used to model
simultaneous translation of several   mRNA molecules while competing for the available ribosomes~\cite{RFM_model_compete_J} (see also~\cite{Algar:cdc2014} for some related ideas).

 It is important to mention that it has been shown in~\cite{reuveni} that the correlation between the production rates based on modeling using RFM and using TASEP over all \emph{S. cerevisiae} endogenous genes is~$0.96$,
 that the RFM agrees well with biological measurements of ribosome densities, and that the RFM predictions correlate well (correlations up to~$0.6$) with protein levels in various organisms (e.g. {\em E. coli, S. pombe, S. cerevisiae}).  More recent results~\cite{Edri2013} show that a certain version of the RFM predicts well the density of RNA polymerases (RNAPs) during transcription.  Given the high levels of bias  related to the state of the art measurements of gene expression and the inherent  noise in intracellular biological processes (see e.g. \cite{Diament2016,Kaern2005}), these are very high correlations that demonstrate the relevance of the RFM in this context.

In this paper, we  analyze the regulation of translation using the RFM.
To do this, we first introduce  two generalizations of the RFM
into a control system.

\subsection{The  Controlled RFM}

\subsubsection{State- and Output-Controllability}

Assume  that  every~$\lambda_i$  can be controlled independently. Thus,
we replace every~$\lambda_i$ in the RFM
by a function~$u_i(t):\R_+\to\R_+$. The set of admissible controls~$\U$
includes all the functions  that are measurable, bounded, and take non-negative values for all~$t\geq 0$.
In the context of   translation, manipulating the~$u_i(t)$s corresponds to dynamically varying
translation factors that regulate the initiation, elongation, and exit rates along the mRNA molecule.
 Note that we may view this as a networked control system:
each state-variable represents an agent, the graph describing the agents interaction is a simple
directed path, and  the~$u_i$s control the strength 
 of the graph edges. 
However,   the dynamics of each agent is nonlinear.

%

The problem we consider is whether it is possible, using the $n+1$ control functions,
to steer~$x$ and~$R$ from any initial condition
to any desired conditions~$x^f \in \Int(C^n)$ and $R^f \in \R_{++}$  in finite time, and if so,
 to determine  appropriate controls.

Of course,  independently controlling all the transition rates
may be difficult
 to do in practice, so we also consider another controlled version of the RFM.

\subsubsection{Output Controllability}
Assume that   a subset of~$m$
 rates $\lambda_{j_1},\dots,\lambda_{j_m}$, with $1\le m\le n+1$,
 can be \emph{jointly} controlled, i.e. all these rates can be replaced by a common, scalar, non-negative control function~$u(t)$.
 This models the case where a single factor jointly controls one or more transition rates.

For example in an RFM with length~$n=3$, assume that the rates~$\lambda_1$ and~$\lambda_2 $ can be replaced by a common, scalar, non-negative control function~$u(t)$. The resulting model is
\begin{align*}
\dot x_1&=\lambda_0 (1-x_1)-u x_1(1-x_2) ,\\
\dot x_2&=u x_1 (1-x_2)-u x_2(1-x_3) ,\\
\dot x_3&=u x_2 (1-x_3)-\lambda_3 x_3 .
\end{align*}

This scenario is biologically relevant  since the exact same codon may appear in multiple places along the transcript, and since the same tRNA species may moreover be involved in the decoding of more than a single codon through wobble pairing. Thus, regulating the abundance of a single tRNA molecule would typically have a simultaneous effect on transition rates at multiple positions along the mRNA transcript.
 In the context of this problem, we are interested in using~$u(t)$ to steer only the production rate~$R$
 to a desired value~$R^f$ in finite time.
Specifically, the problem that we consider is whether it is possible to use~$u$
to steer~$R$ from any initial condition
to any   feasible value and, if so, to determine a suitable control~$u$.
 Of course, the set of feasible values is determined by the other, $n+1-m$ fixed
transition rates.

We show that both control problems described above are \emph{controllable}.
In other words, the control authority is always powerful enough to obtain any feasible desired
density profile and/or production rate.
This is a  primarily   theoretical result.
However, we also show that there exist positive and constant controls
that asymptotically steer the controlled RFM to the desired densities/production rate.
In the problem of controlling all the rates, these constant values are given in a simple and
 closed-form  expression.
In the second control problem, this constant value can be easily found numerically using a simple line search algorithm.

We now discuss the biological relevance of these control problems. 
Understanding and manipulating the
 mRNA translation rate is related to numerous
 biomedical disciplines including 
 human health, evolution, genetics, biotechnology, 
and more~\cite{KimchiSarfaty2007,Kurland1992,Alberts2002,Subramaniam2014,Romanos1992,Moks1987,Binnie1997,Tuller2015,Arava2003}.
Controlling the entire ribosomal density profile, and not  only the translation rate, 
by manipulating  the transition rates 
is also a   fundamental  problem as it is known 
  that the density profile along the mRNA molecule is important for various intracellular phenomena. 
	For example, it was shown that the density and  induced speed of ribosome flow
	along the mRNA affect co-translational folding of the protein.
	If the density and the induced flow speed of the ribosomes is inappropriate 
then  the protein may misfold leading to
 a nonfunctional protein (see, for example,~\cite{KimchiSarfaty2007,Pechmann2013,Kurland1992,Zhang2009}).  
 In addition, it was suggested that the density of ribosomes affects mRNA degradation: a 
higher ribosome density is related to lower efficiency of mRNA degradation and longer half life~\cite{Edri2014,Pedersen2011,Deana2005,Dreyfus2002}. Furthermore,      ribosome density is directly related
 to ribosomal collisions and translation abortion~\cite{Tuller2015,Gorgoni2014,Tuller2010c,Zur2016,Subramaniam2014}: 
a higher density increases the probability of collisions and may lead to abortions and thus the 
production of truncated and potentially deleterious proteins. Finally,  ribosome density 
is strongly correlated  with ribosome allocation: a higher density of ribosomes on the mRNA decreases the pool of free ribosomes, the initiation rate in other mRNA molecules, and thus the organism growth rate and fitness~\cite{Tuller2015,Gorgoni2014,Tuller2010c,Zur2016,Subramaniam2014}.

Our results
suggest
 that these  important issues   can be   addressed  using a combination of  mathematical, computational, and
experimental approaches. 
  Our results also  provide an initial but explicit
solution to the problem  of controlling the augmented profile.  
While  the model  and problems are  relatively simple,
 they may still provide a  reasonable approximation to
the biological  solution in some cases. They may also be used as
a starting  point for  addressing and solving similar problems in
 more comprehensive  models of translation.

The next section describes our main results.
Readers who are not familiar with controllability analysis  may consult
Appendix~A for a quick review of this topic.

\section{Main Results}\label{sec:main}

As noted above,
we  consider two control  problems for the RFM. We now detail their exact mathematical formulation, and then present our main results.

\subsection{Controlling the State and the Output}
Let $\Omega:=C^n \times \R_{+}$. Assume first that all the $n+1$ transition rates can be controlled.
The control is then~$u(t)=\begin{bmatrix}u_0(t),\dots , u_n(t) \end{bmatrix}'$
and the dynamics of the  controlled RFM with output $R(t)$ is described by:
\begin{align}\label{eq:aug_dyn}
\dot{x}_i(t)&=u_{i-1}(t)x_{i-1}(t)(1-x_i(t))-u_i(t) x_i(t)(1-x_{i+1}(t)),\quad i=1,\dots,n, \nonumber \\
R(t) &= u_n(t) x_n(t).
\end{align}
We define the admissible set~$\U$ as the set of measurable and bounded controls taking values in~$\R^{n+1}_+$ for all time~$t$.
\begin{Problem}\label{prob:allc}
					Given arbitrary~$x^s,x^f \in \Int(C^n)$ and~$R^s,R^f \in \R_{++}$, does there always exist  a time~$T\geq 0$
					and a control~$u\in\U$ such that~$x(T,u,x^s)=x^f$ and $R(T,u,R^s)=R^f$? If so, determine such a control.
\end{Problem}
We can now state our
  first main result. Recall that all the proofs  are placed in the Appendix.
\begin{Theorem}\label{thm:con_all_rates_full}
The  controlled RFM~\eqref{eq:aug_dyn} is state- and output-controllable on~$\Int(\Omega)$.
Furthermore, 	for any~$x^f =\begin{bmatrix} x^f_1&\dots&x^f_n \end{bmatrix}'\in \Int(C^n)$ and $R^f\in\R_{++}$,
  define~$v \in \R^{n+1}_{++}$ by
\be\label{eq:l_unique}
v_i : =\frac{R^f}{x^f_{i } (1-x^f_{i+1})}, \quad i=0,\dots,n,
\ee
where $x^f_0:=1$ and $x^f_{n+1}:=0$. Then for any~$x^s\in  C^n$ and any~$ R^s \in \R_{+}$
  applying the constant control~$u(t)\equiv v$ in~\eqref{eq:aug_dyn} yields
\begin{align}\label{eq:foopo}
 \lim_{t\to\infty} x(t,u,x^s)=x^f,\quad
 \lim_{t\to\infty} R(t,u,R^s) = R^f .
\end{align}
\end{Theorem}

This  means that the control is ``powerful'' enough to steer the system, in \emph{finite time}, from any initial {\profile}
to  any desired final {\profile}. It
  also provides a simple closed-form solution for
 a control
that \emph{asymptotically} steers the system to~$x^f$ and $R^f$ from any initial condition. In other words, it   practically
solves  the  control synthesis problem.

  An important property of~$v$ is that it does not depend on the initial values~$x^s$ and $R^s$, but only on the
desired {\profile}~$(x^f,R^f)$. This is important as measuring~$x^s$, that is,
the initial ribosomal  profile along the mRNA,  may be difficult due to the
 current limitations in measuring ribosome densities (see, for example, \cite{Dana2012,Dana2014B,Diament2016}).

\begin{figure}[t]
  \begin{center}
  \includegraphics[height=7cm]{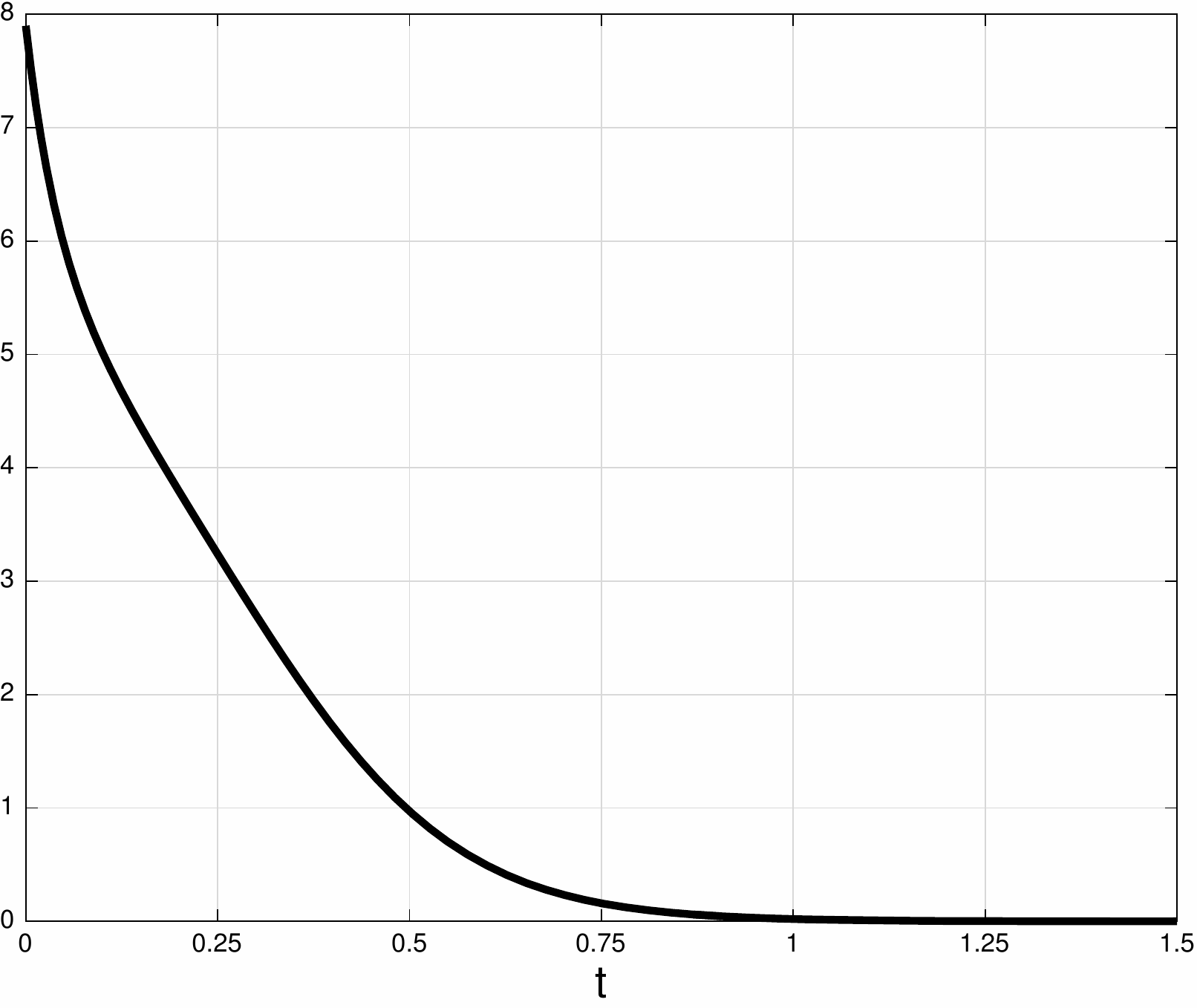}
  \caption{ The error $ |x(t )-x^f |_1+|R(t)-R^f|_1$ as a function of $t$ in Example~\ref{exp:exp1}.} \label{fig:n5_contr}
  \end{center}
\end{figure}

\begin{Example}\label{exp:exp1}
Consider the controlled RFM with dimension~$n=5$.
Suppose that
we would like to steer the ribosomal density profile along the mRNA molecule  to~$\begin{bmatrix} 0.8 & 0.1 & 0.1  & 0.1  & 0.1
 \end{bmatrix}' $, and the production rate to~$1.5$.
The profile here  is motivated by  the fact that  low ribosome abundance at the beginning of the ORF reduces
ribosome ``traffic jams'' that may lead to ribosome drop off.
Setting~$x^f=\begin{bmatrix} 0.8 & 0.1 & 0.1 & 0.1 & 0.1 \end{bmatrix}' $, $R^f=1.5$, and
applying~\eqref{eq:l_unique} yields
\[
v= \begin{bmatrix} 15/2& 25/12& 50/3& 50/3& 50/3& 15  \end{bmatrix}'.
\]
Fig.~\ref{fig:n5_contr} depicts the error~$ |x(t,u,x^s)-x^f|_1+|R(t,u,R^s)-R^f|_1$ (where~$|z|_1$ denotes
 the~$L_1$ norm of the vector~$z$) for  the initial conditions $x^s=\begin{bmatrix}0.5& 0.5& 0.5& 0.5& 0.5\end{bmatrix}'$, $R^s=0.5$, and the control~$u(t)\equiv v$. It may be observed that the error decays at an exponential rate to zero.
Thus, this control  steers  the system arbitrarily close to the desired final density profile $x^f$ and production rate $R^f$.~$\square$
\end{Example}

Example~\ref{exp:exp1} suggests that
the explicit constant control in Theorem~\ref{thm:con_all_rates_full}
provides  a   good practical solution to Problem~\ref{prob:allc}.

\subsection{ Controlling the Output}
Pick an arbitrary set of indexes~$\Theta \subseteq \{0,\dots,n\}$, and let $m:=|\Theta|$.
Replace every~$\lambda_i$, $i\in \Theta$, in the~RFM by a common, scalar control~$u(t)$.
 Pick $c>0$, and assume that $u(t)\in[0,c]$, for all $t\ge 0$, i.e. the set of admissible controls~$\U$ is
the set of measurable scalar functions taking values in~$[0,c]$ for all~$t\geq 0$.
As noted above, this formulation represents a biologically relevant scenario, as
  we assume that several translation rates are controlled by the same control,
  and also that the allowed control action is bounded by the value~$c$.

Our goal is to use the scalar control  to regulate
 the production rate~$R(t)$, i.e.  the output. Of course, not every value of~$R(t)$ is possible, because of the non-regulated, fixed transition rates.
One can in principle define the reachable set of~$R(t)$ based on the fact that the state trajectories evolve on~$C^n$.
 For example, if~$n\notin \Theta$ then~$R(t)=\lambda_n x_n(t)$ implies that one can
define the reachable set as~$[0,\lambda_n]$. However, this definition is not really relevant. Indeed, assume that some rate~$\lambda_k $, with $k\notin\Theta$,
is much smaller than all the other rates and also much smaller than~$c$. Then regardless
of the specific control used it is clear that after some time~$R(t)$ will also be small, as $\lambda_k$
 will be the limiting factor, and so after some time it will become impossible
 to steer the production rate to every   desired value in the set~$[0,\lambda_n]$.

We define  a more meaningful reachable set for the production rate as follows.
Let~$\bar \lambda \in \R_{++}^{n+1-m}$ denote the set of fixed transition rates.
For every time~$T\geq 0$ and every initial  condition~$x_0 \in C^n$,
let~$\Omega(\bar\lambda, \Theta,c, T,x_0)\subset \R_+$ denote the set of production rates
that can be attained at some  time~$t\geq T$ with~$x(0)=x_0$. Define
      the \emph{large-time reachable set} of~$R$ as
      \[
            \Omega(\bar\lambda, \Theta,c,x_0):=\cap_{T\geq 0} \Omega(\bar\lambda,\Theta,c,T,x_0).
      \]
Although the RFM is a nonlinear model, this set  can be characterized explicitly.
To derive this characterization, we introduce more notation.
First,   define a vector~$q\in\R^{n+1}$ by
\[
			q_i:=\begin{cases}
													c,& i \in \Theta,\\
																\lambda_i,& \text{otherwise}.
			\end{cases}
\]
 For example, for $\Theta=\{1,2,n\}$, $q=\begin{bmatrix}
\lambda_0,c,c,\lambda_3,\dots, \lambda_{n-1},c\end{bmatrix}'$.

Also, for~$\ell_0,\dots,\ell_n>0$
define a $(n+2)\times(n+2)$ symmetric, tridiagonal, and componentwise nonnegative  matrix $A=A(\ell_0,\dots, \ell_n)$
by
\be\label{eq:bmatrox}
                A := \begin{bmatrix}
 0 &  \ell_0^{-1/2}   & 0 &0 & \dots &0&0 \\
\ell_0^{-1/2} & 0  & \ell_1^{-1/2}   & 0  & \dots &0&0 \\
 0& \ell_1^{-1/2} & 0 &  \ell_2^{-1/2}    & \dots &0&0 \\
 & &&\vdots \\
 0& 0 & 0 & \dots &\ell_{n-1}^{-1/2}  & 0& \ell_{n }^{-1/2}     \\
 0& 0 & 0 & \dots &0 & \ell_{n }^{-1/2}  & 0
 \end{bmatrix},
\ee
and
 let~$\zeta_{MAX}(A)$
denote the maximal eigenvalue of~$A$.\footnote{It is clear that the eigenvalues are real as~$A$ is symmetric. Since~$A$ is also nonnegative and irreducible the eigenvalues are distinct.} The next result uses the linear-algebraic representation of the steady-state production rate
in the~RFM derived in~\cite{RFM_max}.

\begin{Proposition}\label{prop:ome}
For any~$x_0 \in C^n$,
\be\label{eq:omeghere}
\Omega(\bar\lambda, \Theta,c) =[0,M],
\ee
where~$M:=(\zeta_{MAX}(A(q_0,\dots,q_n)))^{-2}$.
\end{Proposition}

Note that~\eqref{eq:omeghere} implies that $\Omega(\bar\lambda, \Theta,c)$ does not depend on $x_0$, but only on the vector~$q$.

\begin{Remark}\label{remark:cinf}
Denote the indexes in  $\Theta$ by $j_1,\dots,j_m$. Consider the case~$c\to \infty$. Then $c^{-1/2} \to 0$, so the largest eigenvalue  of the matrix
$
A(q_0,\dots, q_n)
$
 tends to
\[
\max\{ \zeta_{MAX}(Q_0),\dots, \zeta_{MAX}(Q_{m})\},
\]
where
\begin{align}\label{eq:Q_mat}
Q_0&:=A(\lambda_0,\dots,\lambda_{j_1-1}), \nonumber \\
Q_k&:=A(\lambda_{j_{k}+1},\dots,\lambda_{j_{k+1}-1}), \quad k=1,\dots,m-1, \nonumber  \\
Q_{m }&:=A(\lambda_{j_{m}+1},\dots,\lambda_{n}),
\end{align}
with $\zeta_{MAX}(B):=0$ if $B$ is an empty matrix. Thus, in this case
\be\label{eq:minrr}
 M= \min\{  (\zeta_{MAX}(Q_0))^{-2},\dots, (\zeta_{MAX}(Q_{m}))^{-2}   \},
\ee
where~$0^{-2}$ is defined as~$\infty$.
In other words, when the maximal control value
of the controlled transition rates  goes to infinity, the maximal possible steady-state production rate will be the minimum of the steady-state production rates of~several RFMs: the first with  rates $\lambda_0,\dots,\lambda_{j_1-1}$,
the second with   rates
 $\lambda_{j_{1}+1},\dots,\lambda_{j_{2}-1}$, and so on, with the last RFM
with   rates $\lambda_{j_{m}+1},\dots,\lambda_{n}$.
This demonstrates  how in this
 case the other, fixed rates, being the limiting factors, determine the feasible
  set for  the production rate.
\end{Remark}

From the biological point of view this means
 that if the transition rates along some regions of
the mRNA are  very high (and thus not rate limiting)   the production rate will depend
  only on the transition rates before and after this region, as these  include the rate limiting factor.
Also, the large-time reachable set for the production rate
will be constrained by the rate limiting transition rates.

\begin{figure}[t]
 \begin{center}
 \includegraphics[height=7cm]{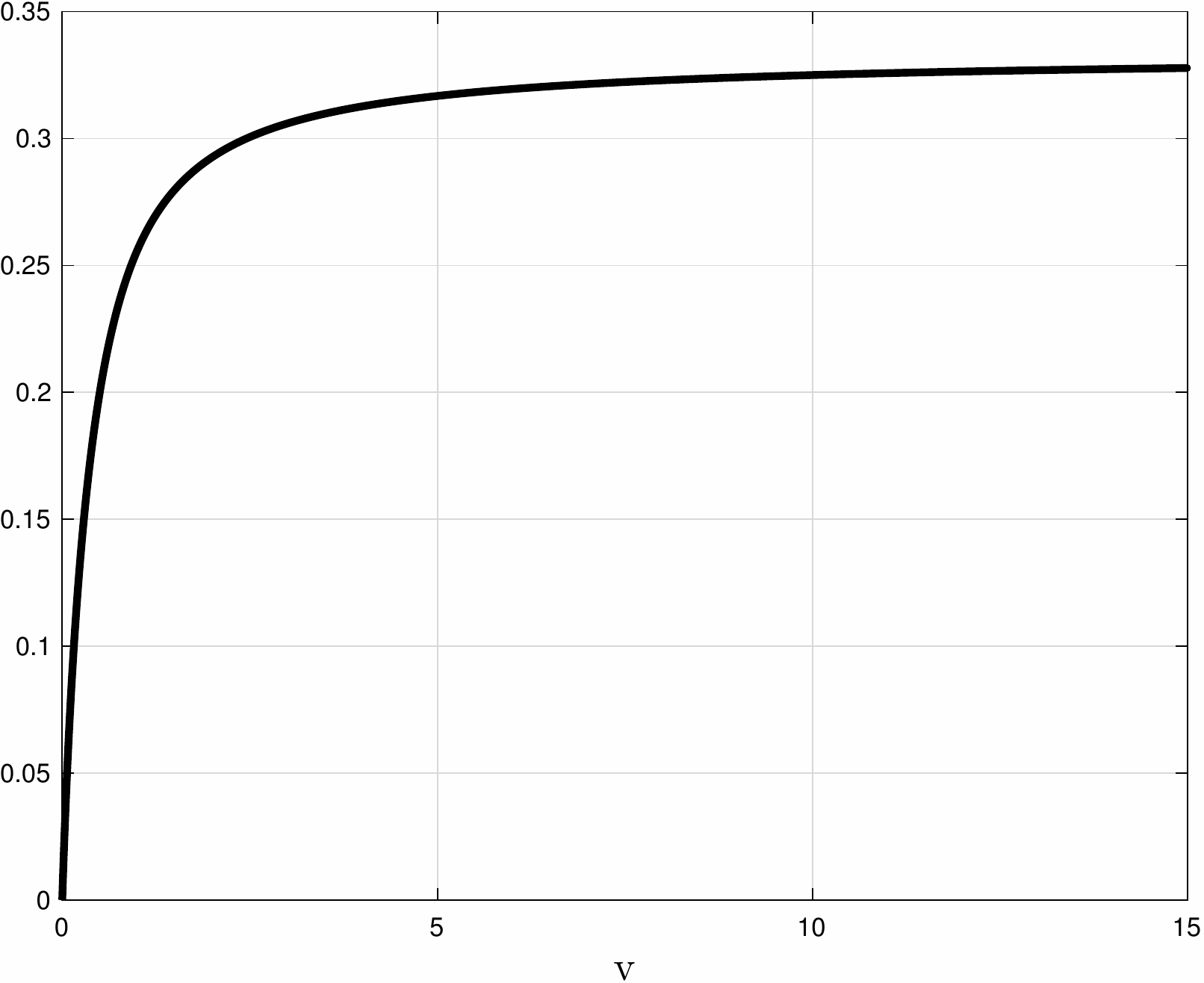}
 \caption{Maximal  steady-state production rate~$(\zeta_{MAX}(A(1,1/2,v,3,v,1/2)))^{-2}$
 for~$v \in [0,15]$.   } \label{fig:rfm_n5_m2}
\end{center}
\end{figure}

\begin{Example}\label{exa:addnt_n5}
Consider a controlled RFM with length~$n=5$,~$\Theta=\{2,4\}$, and fixed rates
\be\label{eq:addrae}
\lambda_0=1, \; \lambda_1=1/2, \; \lambda_3=3,\; \lambda_5=1/2.
\ee
In other words, $\lambda_2$ and $\lambda_4$ are both replaced by the scalar control~$u(t)$. Suppose that the  admissible set~$\U$ is the set of functions  taking values in~$[0,c]$, with~$c=15$.
Fig.~\ref{fig:rfm_n5_m2} depicts~$(\zeta_{MAX}(A(1,1/2,v,3,v,1/2)))^{-2}$ for~$v \in [0,15]$. It may be seen that this  is a strictly increasing function of~$v$. A calculation yields (all numbers are to four digit accuracy):
\[
(\zeta_{MAX}(A(1,1/2,15,3,15,1/2)))^{-2}=0.3278,
\]
so~$\Omega =[0,0.3278]$.
																								
Note that if we take~$c\to\infty$ then~\eqref{eq:Q_mat} yields
\begin{align*}
Q_0&=\begin{bmatrix}   0 & 1 &0 \\
           1& 0& (1/2)^{-1/2}   \\0& (1/2)^{-1/2} &0 \end{bmatrix}, \\
Q_1&=  \begin{bmatrix}   0 & 3^{-1/2} \\
                                          3^{-1/2} &0 \end{bmatrix}, \\
Q_2&=\begin{bmatrix}   0 & (1/2)^{-1/2} \\
                                          (1/2)^{-1/2} &0 \end{bmatrix},
\end{align*}
and so~\eqref{eq:minrr} yields
\begin{align*}
\min\{  (\zeta_{MAX}(Q_0))^{-2}  ,
                                         (\zeta_{MAX}( Q_1 ) )^{-2},
                                          (\zeta_{MAX}(Q_2  ) )^{-2}   \}     &=\min\{1/3 , 3,1/2\}\\&=1/3.
\end{align*}
~$\square$
\end{Example}

The next result considers controlling the output to a desired
value in~$\Omega(\bar\lambda,\Theta,c)$.
\begin{Proposition}\label{prop:asyconsingle}
The controlled RFM with one or more rates  replaced by a common scalar control function~$u(t)$ is output-controllable  in~$\Int( \Omega(\bar\lambda,\Theta,c) )$.
Furthermore, for any~$R^f \in \Int(\Omega(\bar\lambda,\Theta,c) )$
there exists a value~$v\in[0,c]$ such that the constant control~$u(t)\equiv  v$
yields~$\lim_{t\to\infty}R(t)=R^f$.
\end{Proposition}
This  means that jointly regulating one or more transition rates with a common scalar control function~$u(t)$ is still
   ``powerful" enough to steer the production rate  from any initial value
to  any desired final  value~$R^f \in \Int(\Omega )$ in finite time.
Furthermore,  the controlled RFM is
 asymptotically controllable in~$ \Omega $, even
when~$\U$ is restricted     to constant controls only.
Since~$\zeta_{MAX}(A(\ell_0,\dots,\ell_n))$ is  a strictly decreasing  function of every~$\ell_i$,
finding the constant value~$v$ that asymptotically  steers the system to a desired value~$R^f\in\Int(\Omega ) $
  can be easily solved numerically using a simple line search. The next example demonstrates this.

\begin{Example}
                                Consider again the controlled RFM in Example~\ref{exa:addnt_n5}.
                                  Recall  that the  admissible set~$\U$ is the set of
                                 functions  taking values in~$[0,c]$, with~$c=15$.
																We already know that in this case~$\Omega =[0,0.3278]$.
                        Assume that our goal is to asymptotically steer the production rate to, say,~$R^f=0.3$.
A simple line search shows that the corresponding constant control value
is~$v=2.4534$ (see also Fig.~\ref{fig:rfm_n5_m2}).~$\square$

\end{Example}

\subsection{Sensitivity analysis}\label{subsec:sense}

In practice, the applied controls are never exactly equal to the desired
 values and therefore it
is important to understand the effect of small perturbations in the control values on the desired {\profile}. Since we are basically considering
constant controls, it is enough to study the sensitivity of the steady-state density profile of the~RFM to small changes in the~$\lambda_i$s.
(The sensitivity of the steady-state production rate~$R$ with respect to the~$\lambda_i$s has been studied in~\cite{RFM_sense}.)
\begin{Proposition}\label{prop:sens_u}
Consider the RFM with dimension $n$, and let~$e:=\begin{bmatrix} e_1&\dots&e_n\end{bmatrix}'$ denote the corresponding
equilibrium point in~$\Int(C^n)$. Pick an index~$i\in\{0,\dots,n\}$. Then~$\frac{\partial}{\partial \lambda_i} e_k$ exists for all~$k$, and
\begin{align}
&\frac{\partial}{\partial \lambda_i} e_k < 0, \quad \text{for all } k\leq i, \nonumber \\
&\frac{\partial}{\partial \lambda_i} e_k > 0, \quad \text{for all } k>i.
\end{align}
\end{Proposition}

Thus, increasing~$\lambda_i$  decreases [increases]  the steady-state densities in sites~$1,\dots,i$
[sites~$i+1,\dots,n$]. This is   reasonable,  as increasing~$\lambda_i$ increases the
transition rate from site~$i$ to site~$i+1$ (see also~\cite{RFM_model_compete_J} for some related considerations).

\begin{Example}\label{exp:exp_sense}
Recall from Example~\ref{exp:exp1} that for the RFM
 with~$n=5$ the control
\[
	u(t)\equiv      \begin{bmatrix} 15/2 & 25/12& 50/3& 50/3& 50/3& 15  \end{bmatrix}',
\]
yields the steady-state {\profile}:
\be\label{eq:profile}
				\begin{bmatrix} e& R\end{bmatrix}'=\begin{bmatrix} 0.8 & 0.1 & 0.1 & 0.1 & 0.1 &  1.5\end{bmatrix}' .
\ee
	Let~$\tilde u(t)			 \equiv    \begin{bmatrix} 15/2 & 25/12& (50/3)+\varepsilon & 50/3& 50/3& 15\end{bmatrix}' $,
with~$\varepsilon:=0.2$ i.e. the same transition rates as before, but with~$\varepsilon$
added to~$\lambda_2$.
Using~\eqref{eq:ep} shows  that~$\tilde u$ yields the  steady-state {\profile}
\[
				\begin{bmatrix} \tilde e& \tilde R\end{bmatrix}'=\begin{bmatrix}
			 0.7998&  0.0989& 0.1001& 0.1001& 0.1001&  1.5013
\end{bmatrix}'
\]
(all numbers are to four digit accuracy).
Comparing this to~\eqref{eq:profile} shows that the steady-state values at sites~$1,2$ decreased, and those at
sites~$3,4,5$ increased.~$\square$
\end{Example}

\section{Discussion}

Regulating the ribosomal density profile along the mRNA molecule, and not only the protein production rate,
is an important problem in evolutionary biology, biotechnology, and synthetic biology
 because  this density profile     affects   various fundamental intracellular processes including mRNA degradation, protein folding, ribosomal allocation and abortion, and more (see, for example, \cite{Zhang2009,Kurland1992,Edri2014,KimchiSarfaty2007,Pechmann2013,Tuller2015}).
It seems that there are still considerable gaps in our understanding of how the density profile is regulated,
 and how it can be  re-engineered.
In this paper, we  addressed  this issue by analyzing
 a mathematical model for ribosome flow, the RFM,
using  tools from nonlinear  control theory.

Our results  indicate  that if we are able to control all  the transition rates  along the  different parts of the mRNA then
we can   steer the system to any desired ribosomal density profile,
and we provide a closed-form expression for
a constant control vector that achieves this asymptotically.

Also, jointly
controlling one or more transition rates using a common
scalar control allows to steer
the  protein production rate to any desired value within a feasible range that is determined by the other, fixed transition rates. A simple line search algorithm can be used to derive
a constant
control value that achieves this asymptotically.
This  case models scenarios where for example the abundance of
a  specific loaded tRNA molecule is  regulated.
Indeed,   regulating the abundance of a certain tRNA molecule  should simultaneously
affect the translation rate at all the positions along the mRNA with corresponding codons.
 Typically, a certain codon may repeat at  dozens, or even hundreds of  locations  along one  mRNA molecule.

Our results are based on the RFM that, as any mathematical model, is a simplification of
(the biological) reality. For example, the RFM does not encapsulate some of the
complex interactions between the transcript features and translation (see, e.g., \cite{TullerGB2011,Sabi2015,Tuller2015}).
Nevertheless, using the RFM
 allows one to pose the controllability   and control synthesis problems
in a well-structured way, and study them
 rigorously using tools from systems and control theory.

We believe that our analytical results
may lead to new biological  insights    and  suggest novel and  interesting  biological experiments.
For example,
it has been  suggested that a  higher ribosome density contributes to a higher mRNA half life in {\em S. cerevisiae}~\cite{Edri2014}. However, it is difficult  to determine
if the correlation  is due  to a larger abundance of ribosomes along the entire coding region or maybe only the ribosome density at the 5'end of the coding region is relevant. It is also possible that this relation is due
 to a  higher number of pre-initiation complexes at the 5'UTR (that contribute to a  higher initiation rate). Specifically, it is possible that only higher pre-initiation density or ribosome density at the 5'end is important since in some cases the degradation starts from this region.  Both factors  are expected to correlate with higher ribosome density along the entire coding region, and a natural
question is how  can we design an experiment that can separate between the two possible explanations?

The results reported here suggest that we can  design
 a synthetic library (that can be studied in-vitro and/or in-vivo) with   different strains that
have different initiation rates, but identical ribosome densities along the coding regions,
 or strains with different
levels of ribosome densities at the first codons (or any other segment) of the coding regions, but similar ribosome densities in the rest of the coding region.
Using such libraries may help in understanding exactly which factor contributes to  the higher
  mRNA half life.

Regulating transition rates can also affect the folding of the protein. Indeed, it was suggest in~\cite{nissley2016accurate} that synonymous codons substitutions, that  change the corresponding transition rates, may switch some protein domains between post-translationally and co-translationally folding.

We believe that the results reported in this study may also contribute   towards a  better understanding of the molecular evolution of translation. Since usually a change in a transition rate is related to a  mutation/change in the mRNA codons composition, obtaining a desired ribosomal density profile and production rate involves introducing changes in the nucleotide composition of the transcript. Thus,  an important future study should combine  controllability analysis  with models of molecular evolution.


Other topics for further research  include the following. First, from the biological point of view
 a  relevant scenario is when some of
the transition rates can be   controlled, but  each rate
can take values in   a discrete  set of possible values only.
  Indeed, the admissible rates
are limited by factors such as
  the concentrations of initiation and elongation factors, and the biophysical properties of the ribosome, mRNA, and translation factors.
In this case, it is clear that we cannot obtain any desired density
profile, and an interesting problem may be to determine the rate values that yield the
``best'' approximation for a given  desired profile. This requires
  a biologically relevant definition of this best approximation, i.e., a
  measure of distance between two density
	profiles  that  is biologically relevant.

Second, as noted above, the RFM is the mean-field approximation of~TASEP.
Our results naturally raise the question of whether~TASEP is controllable (in some stochastic sense).
It is also interesting to examine if the analytical results obtained for the RFM can be used
to synthesize suitable hopping rates for the  stochastic~TASEP model.
In other words, suppose that we are
 given a desired profile~$P$ for the~RFM, and determine  the corresponding     constant rates~$v_i$s
 using~\eqref{eq:l_unique}. Does using these  rates (perhaps after some normalization)   as the~TASEP hopping rates
yield the steady-state profile~$P$ in TASEP as well?


Finally, TASEP has been used to model and analyze many other applications, for example, traffic flow.
The~RFM can also be used to study these applications, and controllability and control synthesis
may be important here as well.
For example, a natural question is can the density along a traffic lane be steered to any
arbitrary profile by regulating speed  signs along   different sections of the lane?

\section*{Acknowledgments}
We thank Pablo Iglesias for helpful comments.
We are grateful to the anonymous reviewers and the~AE for  comments that helped us to greatly improve this paper.


\appendix
\counterwithin{Definition}{section}
\counterwithin{Theorem}{section}
\counterwithin{Example}{section}
\counterwithin{Fact}{section}

\section*{Appendix A: review of controllability}\label{sec:cont}

Controllability is a fundamental property of control systems,
but it is not necessarily  well-known  outside of
the systems and control community.
 For the sake of completeness, we briefly  review this topic here.
    For more details, see e.g.~\cite{sontag_book}.

Consider the control system
\begin{align}\label{eq:contsys}
\dot x &=f(x,u), \nonumber \\
y &=h(x,u),
\end{align}
where~$x:\R_+ \to  \R^n$ is the state vector,~$u:\R_+ \to \R^m$ is the control, and~$y:\R_+ \to  \R^k$ is the output.
Let~$\U$ denote the set of admissible controls.
Assume that the trajectories of this system evolve on a state space~$\Omega \subseteq \R^n$.
 Given an initial condition~$a\in \Omega$
and a desired final condition~$b\in\Omega$, a natural control problem is: find a time~$T\geq 0$,
and an admissible control~$u:[0,T]\to\R^m$ such that
\[
x(T,u,a )=b.
\]
In other words,~$u$ steers the system from~$a$ to~$b$ in time~$T$. Of course, such a control may not always  exist.
This leads to the following definition.
\begin{Definition}
The system~\eqref{eq:contsys} is said to be \emph{state-controllable} on~$\Omega$
if for any~$a,b\in \Omega$ there exist a time~$T\geq 0$, and a control~$u\in \U$ such that
$x(T,u, a )=b.$
\end{Definition}

Sometimes it is enough to steer  only the output to a desired condition. This leads to the following definition.
\begin{Definition}
The system~\eqref{eq:contsys} is said to be \emph{output-controllable} on some set~$\Psi\subseteq \R^k$
if for any~$p,q\in \Psi$ there exist a time~$T\geq 0$, and a control~$u\in \U$ that steers
 the output from $y(0)=p$  to $y(T)=q$.
\end{Definition}

Controllability is thus a theoretical property, but it is important in many applications, as
it implies that the problem of determining a suitable control, i.e. the \emph{control synthesis problem}, always admits a solution.
From here on we focus on state-controllability. The notions  for output-controlability are analogous.

Another useful notion, that is weaker than controllability,  is called asymptotic controllability.
\begin{Definition}
System~\eqref{eq:contsys} is said to be \emph{asymptotically state-controllable} on~$\Omega$
if for any~$a,b\in \Omega$ there exists  a control~$u\in \U$ such that
\[
\lim_{t\to\infty}x(t,u,a)=b.
\]
\end{Definition}
Note that this implies that for any neighborhood~$V$ of~$b$, there exists a time~$T_s\geq 0$,
and a control~$u_s\in\U$ such that~$x(T_s,u_s,a) \in V$.

For nonlinear control systems, analyzing controllability or asymptotic controllability
is not  trivial. There exists a weaker theoretical
notion
that can be analyzed effectively using Lie-algebraic techniques.
For~$a\in\Omega$, define the \emph{reachable set from~$a$}  by
\[
			RS(a):=\{ x(t,u, a): t\geq 0,\; u\in \U\}.
\]
In other words,~$RS(a)$ is the set of all states that can be reached at some time~$t\geq 0$
starting from~$x(0)=a$.
The system~\eqref{eq:contsys} is said to be \emph{accessible} from~$a$
if the set~$RS(a)$ has a non empty interior. In other words, the control is powerful  enough to
allow steering the trajectories  emanating from~$a$ to a ``full set'' of directions.


\begin{Example}
Consider the scalar system~$\dot x=u$, with~$\Omega=\R$. Let~$\U$ be the set of measurable functions
taking non negative values for all time~$t$. Pick~$a\in\Omega$. Then~$RS(a)=[a,\infty)$, so the systen
is accessible  from~$a$.
   However, the system is not controllable on~$\Omega$, as there does not exist any control~$u\in\U$ that steers~$a$ to a point~$b$
	with~$b<a$.~$\square$
	\end{Example}

Our results for the controlled RFM are based on proving
 that it is asymptotically state-controllable,
using \emph{constant} controls, and combining this with
 a Lie-algebraic sufficient condition for accessibility
  to deduce state-controllability.

To describe a sufficient condition for accessibility,
consider the \emph{control affine system}:
\be\label{eq:affine}
				\dot x =f(x)+\sum_{i=1}^m g_i(x)u_i,
\ee
and assume that~$0 \in \U$. For two vector fields~$f,g:\R^n\to\R^n$, let~$[f,g]:=\frac{\partial g}{\partial x}f-\frac{\partial f}{\partial x}g $.
This is another  vector field  called the \emph{Lie-bracket of~$f$ and~$g$}.
For example, if~$f(x)=Ax$ and~$g(x)=Bx$ then~$[f,g](x)=(BA-AB)x$.
It is useful to introduce a notation for iterated Lie brackets.
These can be defined inductively by letting  $\ad_f^0 g := g$,  $\ad_f^1 g := [f,g]$, and
$\ad_f^k g : = [f, \ad_f^{k-1} g]$ for any integer~$k\geq 1 $.

The \emph{Lie algebra}~$A_{LA}$ associated with~\eqref{eq:affine}
is the  linear subspace that is generated by~$\{f,g_1,\dots,g_m\}$
and is closed under the Lie bracket operation. Let
\[
            A_{LA}(x_0):=\{ p(x_0):p\in A_{LA}  \}.
\]
Roughly speaking, it can be shown that if small-time
 solutions of~\eqref{eq:affine} emanating from a point~$x_0$  and
 corresponding to  piecewise constant controls
``cover''   a $k$-dimensional  set, with~$k\le n$,  then~$A_{LA}(x_0)=\R^k$.
This yields the following   sufficient condition for accessibility.
\begin{Theorem}\label{thm:acc_LA}
If~$A_{LA}(x_0)=\R^n$ at some point~$x_0$ then~\eqref{eq:affine} is accessible from~$x_0$.
\end{Theorem}

The next result applies Theorem~\ref{thm:acc_LA}
to analyze  accessibility in the~RFM  when   either
the entry rate or exit rate is replaced by  a control.

\begin{Fact}\label{fact:acc}
Consider the~$n$-dimensional RFM with a single rate~$\lambda_i$ replaced by a scalar control~$u(t)$.
If~$i=0$ or~$i=n$ then the control  system is accessible from any point~$x\in \Int(C^n)$.
\end{Fact}

{\sl Proof of  Fact~\ref{fact:acc}.}
Consider the controlled RFM obtained by replacing~$\lambda_0$ by~$u(
t)$, leaving the other rates as strictly
positive constants. Let~$z_0(x):=\lambda_0 (1-x_1)$, $z_j(x):=\lambda_j x_j(1-x_{j+1})$, for~$j=1,\dots,n-1$,
and~$z_n(x):=\lambda_n x_n$.
The controlled  RFM satisfies:
\be\label{eq:conrfmu}
			\dot x=f(x)+g(x)u,
\ee
where~$f:=\begin{bmatrix} -z_1 & z_1-z_2&z_2-z_3   &\dots&  z_{n-1}-z_n \end{bmatrix}'$,
 and~$g:=\begin{bmatrix} 1-x_1 &0&\dots&0\end{bmatrix}'$.
Let~$p^k(x):=(\ad_f^k g) (x)$.
A calculation shows  that for all $k\in\{0,\dots,n-1\}$,
\[
p^k=\begin{bmatrix} p^k_1&\dots &p^k_k&p^k_{k+1}&0&\dots&0  \end{bmatrix}',
\]
with~$p^k_{k+1}= (-1)^k \prod_{j=1}^{k+1} (1-x_j) \prod_{\ell=1}^{k}\lambda_\ell$.
Note that~$ p^k_{k+1} \not = 0 $ for all~$x\in\Int(C^n)$, so
 the~$n$ vector fields $p^0,\dots,p^{n-1}$ are linearly
 independent, and thus   span~$\R^n$. Thus, the controlled RFM  is accessible from any~$x\in\Int(C^n)$.

Now consider the case where~$\lambda_n$
 is replaced by a control~$u(t)$.
 For~$j=1,\dots,n$, let~$q_j(t):=1-x_{n+1-j}(t)$.
Then
\begin{align*}
										\dot q_1&= (1-q_1)u -\lambda_{n-1}q_1(1-q_2),      \\
									\dot q_2&= \lambda_{n-1}q_1(1-q_2) -\lambda_{n-2}q_2(1-q_3)  ,    \\
									&\vdots\\
									\dot q_n&= \lambda_{1}q_{n-1}(1-q_n) -\lambda_{0}q_n    .
\end{align*}
This is a controlled RFM with the initiation rate replaced by a control~$u(t)$.
It follows from the analysis   above that this control system is accessible in~$\Int(C^n)$, and
this completes the proof.~\IEEEQED

Another  sufficient condition for accessibility   is based on \emph{linearizing} the control system around an equilibrium point.
For our purposes, it is enough to state this condition
for the control affine system~\eqref{eq:affine}
with~$m=1$, i.e. the    system:
\be\label{eq:affine1}
				\dot x =f(x)+ g (x)u .
\ee
\begin{Theorem}\label{thm:linc} \cite[Ch.~3]{sontag_book}
  Suppose that~$f( e)=0$ and that~$0\in \Int\U$.
Consider the linear control system
\[
                \dot z= Az+ u b,
\]
where~$A:=\frac{\partial f}{\partial x}( e)$ and~$b:=g( e)$.
If the~$n\times n$ matrix~$ \begin{bmatrix} b & Ab &\dots A^{n-1}b \end{bmatrix}$ is invertible
then~\eqref{eq:affine1}   is accessible from some neighborhood of~$ e$.\footnote{In fact, the condition above guarantees a  stronger property, called  first-order local controllability, but for our purposes
the more restricted statement in Theorem~\ref{thm:linc} is enough.}
\end{Theorem}

\begin{Example}\label{exa:con_rfm_not}
Consider the RFM with~$n=2$, i.e.
\begin{align}
            \dot x_1&=\lambda_0(1-x_1)-\lambda_1x_1(1-x_2) ,\\
            \dot x_2&=\lambda_1 x_1(1-x_2)-\lambda_2 x_2,\nonumber
\end{align}
with~$\lambda_i>0$.
The equilibrium point~$e$ of this system satisfies~$\lambda_0(1-e_1)=\lambda_1e_1(1-e_2)=\lambda_2 e_2$.
Suppose now that we can control the transition rate from site~$1$ to site~$2$.
 To study  state-controllability in the neighborhood
 of~$e$, consider the
control system
\begin{align}\label{eq:rfm2n}
            \dot x_1&=\lambda_0(1-x_1)-(\lambda_1+u) x_1(1-x_2) ,\\
            \dot x_2&=(\lambda_1+u) x_1(1-x_2)-\lambda_2 x_2,\nonumber
\end{align}
where~$\U$ is the set of measurable functions taking values in~$[-\varepsilon,\varepsilon]$ for some sufficiently
 small~$\varepsilon>0$. This system is in the form~\eqref{eq:affine1} with
$
            f(x) = \begin{bmatrix}   \lambda_0(1-x_1)-\lambda_1x_1(1-x_2) &   \lambda_1
x_1(1-x_2)-\lambda_2 x_2      \end{bmatrix}'$, and
$            g(x) =x_1(1-x_2) \begin{bmatrix}    -1 &  1      \end{bmatrix}'$.
Note that~$f(e)=0$. To apply Theorem~\ref{thm:linc}, calculate
$A=\begin{bmatrix}    -\lambda_0 - \lambda_1 (1-e_2) &  \lambda_1 e_1 \\
 \lambda_1(1-e_2)&-\lambda_1e_1-\lambda_2    \end{bmatrix}$,  $b= e_1(1-e_2) \begin{bmatrix}   -1&1     \end{bmatrix}'$,  and
\[
\begin{bmatrix} b& Ab \end{bmatrix}=e_1(1-e_2)  \begin{bmatrix} -1&  \lambda_0 + \lambda_1 (1-e_2)+\lambda_1e_1 \\
1 & - \lambda_1(1-e_2) -\lambda_1e_1-\lambda_2  \end{bmatrix}.
\]
Note that~$\det \left (\begin{bmatrix} b& Ab \end{bmatrix} \right )=e_1^2(1-e_2) ^2 (\lambda_2-\lambda_0)  $.
Since~$e\in\Int(C^2)$,
Theorem~\ref{thm:linc} implies that if~$\lambda_0\not = \lambda_2$
then~\eqref{eq:rfm2n} is  accessible
 in a neighborhood of~$e$.

Now consider~\eqref{eq:rfm2n} with~$\lambda_0 = \lambda_2$.
Then~$z:=x_1+x_2$ satisfies
\[
            \dot z=\lambda_0(1-z) .
\]
Thus, any trajectory with~$x_1(0)+x_2(0)=1$ satisfies~$x_1(t)+x_2(t)\equiv 1$ for any control~$u$, and this implies that
in this case~\eqref{eq:rfm2n} is not accessible and not state-controllable on~$C^2$.

Summarizing, in this case the condition in Theorem~\ref{thm:linc}
allows us to completely analyze the accessibility of~\eqref{eq:rfm2n}.~$\square$
\end{Example}

This example may suggest that   accessibility is lost when one of the \emph{internal}
 (or elongation) rates~$\lambda_i$, $i\in\{1,\dots,n-1\}$,
is replaced by a control, at least for some values of the other rates.
 However, the next example shows that is not necessarily  true.

\begin{Example}\label{exa:access3}
Consider the RFM with~$n=3$, i.e.
\begin{align*}
            \dot x_1&=\lambda_0(1-x_1)-\lambda_1x_1(1-x_2) ,\\
            \dot x_2&=\lambda_1 x_1(1-x_2)-\lambda_2 x_2(1-x_3),\\
						\dot x_3&=\lambda_2 x_2(1-x_3)-\lambda_3 x_3 ,
\end{align*}
with~$\lambda_i>0$.
Suppose   that we can control the transition rate from site~$1$ to site~$2$,
so we  consider the
control system:
\begin{align}\label{eq:rfm33}
            \dot x_1&=\lambda_0(1-x_1)-x_1(1-x_2) u,\nonumber \\
            \dot x_2&=  x_1(1-x_2)u-\lambda_2 x_2(1-x_3),\nonumber \\
						\dot x_3&=\lambda_2 x_2(1-x_3)-\lambda_3 x_3 .
\end{align}
We may ignore the term~$x_1(1-x_2)$
 multiplying~$u$,  as it is strictly
positive for all~$x\in\Int(C^3)$. Thus, the control system
  is in the form~\eqref{eq:affine1} with
$
            f(x) = \begin{bmatrix}   \lambda_0(1-x_1)  &
 -\lambda_2 x_2   (1-x_3)&  \lambda_2 x_2   (1-x_3)-\lambda_3 x_3  \end{bmatrix}'$, and
$            g(x) =  \begin{bmatrix}    -1 &  1    &0  \end{bmatrix}'$.
A calculation yields
\[
v^1: =  [f,g]=\begin{bmatrix} -\lambda_0& \lambda_2(1-x_3) & -\lambda_2(1-x_3)   \end{bmatrix}' ,
\]
\[v^2: = [  [f,[f,g]], [f,g] ] =
\begin{bmatrix} 0&  \lambda_2^2 ( \lambda_3(2-x_3) -\lambda_2(1-x_3)^2)
 & \lambda_2^2 ( \lambda_2(1-x_3)^2 -\lambda_3 ) \end{bmatrix}',
\]
\[
v^3:= [ v^2,v^1]=
\begin{bmatrix} 0&  \lambda_2^3 ( \lambda_3 x_3 +\lambda_2(1-x_3)^2)
 & -\lambda_2^3 ( \lambda_2(1-x_3)^2 +\lambda_3 ) \end{bmatrix}',
\]
and
\begin{align*}
				\det \left ( \begin{bmatrix}  v^1& v^2 &v^3   \end{bmatrix} \right )
				&=2\lambda_0 \lambda_2^5 \lambda_3^2 (1-x_3).
\end{align*}
Since this is different from zero  for all~$x\in\Int(C^3)$, we conclude that~\eqref{eq:rfm33}
 is accessible from every~$x\in\Int(C^3)$.~$\square$
\end{Example}

\section*{Appendix B: Proofs}

{\sl Proof of Theorem~\ref{thm:con_all_rates_full}.}

The proof of~\eqref{eq:foopo}
 follows immediately from Remark~\ref{rem:Linvert}.
Indeed, using the constant control~$u(t)\equiv v$ amounts to setting the desired density profile~$x^f$ as the steady-state densities of the dynamics, and $R^f$ as the steady-state production rate. Since this steady-state   is globally asymptotically stable on~$\Int(\Omega)$, this implies~\eqref{eq:foopo}.

We now turn to prove that the system is state- and output-controllable, that is, that we can steer the system to the desired {\profile} $x^f \in \Int(C^n), R^f\in \R_{++}$ in  finite time.
We begin by defining a new control system obtained by
  replacing~$\lambda_i$, $i\in\{0,\dots,n-1\}$,  in the~RFM~\eqref{eq:rfm}
by a control function~$u_i(t):\R_+\to\R_+$ (but leaving~$\lambda_n$ as a constant rate).
This yields
\be\label{eq:rfm_control_affine}
\dot{x}=g_0(x)+\sum_{i=1}^n u_{i-1} g_i(x),
\ee
where
$g_0(x)  :=\begin{bmatrix} 0 & \dots & 0 & -\lambda_n x_n \end{bmatrix}'$,
$g_1(x)  :=\begin{bmatrix} 1-x_1 & 0 & \dots & 0 \end{bmatrix}'$,
and for any~$j\geq 2 $, $g_j(x)$ contains the value $-x_{j-1}(1-x_j)$ in its $(j-1)$'th coordinate, the value $x_{j-1}(1-x_j)$ in its $j$'th coordinate, and the value $0$ otherwise.
For example,
for  $n=4$:
\begin{align*}
g_0(x) &= \begin{bmatrix} 0 & 0 & 0 & -\lambda_4 x_4 \end{bmatrix}', \\
g_1(x) &= \begin{bmatrix} 1-x_1 & 0 & 0 & 0  \end{bmatrix}', \\
g_2(x) &= \begin{bmatrix} -x_1(1-x_2) & x_1(1-x_2) & 0 & 0  \end{bmatrix}', \\
g_3(x) &= \begin{bmatrix} 0 & -x_2(1-x_3) & x_2(1-x_3) & 0  \end{bmatrix}', \\
g_4(x) &= \begin{bmatrix} 0 & 0 & -x_3(1-x_4) & x_3(1-x_4)  \end{bmatrix}'.
\end{align*}
Pick~$z\in\R^n$.
Then
it is straightforward to show that
\[
z=\sum_{i=1}^n \alpha_i g_i(x^f),
\]
where
\[
\alpha_i:=\frac{\sum_{k=i}^n z_k}{x^f_{i-1}(1-x^f_i)},
\]
with $x_0^f:=1$. Since $x^f \in\Int(C^n)$, $\alpha_i$ is well-defined for all $i=1,\dots,n$.
We conclude that
the vector fields~$g_1(x^f),\dots,g_n(x^f)$ span~$\R^n$. This
  implies, by known accessibility results (see, e.g.~\cite[Ch.~4]{sontag_book}),
that there exists a set $V=V(x^f) \subseteq \Int(C^n)$, that   has a  nonempty interior in~$\R^n$,
and such that every~$p \in V$ can be steered to~$x^f$ in \emph{finite time}.
   Fix arbitrary~$q \in \Int(V)$ and~$x^s \in C^n$.
   We already know that there exist constant controls~$u_0,\dots,u_{n}$
	such that~$\lim_{t\to\infty}x(t,u,x^s)=q$, $\lim_{t\to\infty}R(t,u,x^s)=R^f$.
	Therefore there exists a time~$\tau>0$ such that~$x(\tau,u,x^s) \in V$.
	We also know that we can keep~$u_n$ at this constant value, and
	find a time-varying control~$w(t)=\begin{bmatrix} w_0(t),\dots,  w_{n-1}(t), w_n(t) \end{bmatrix}$,~$t \in [\tau,T]$, with~$w_n(t)\equiv u_n$, such that
the  time-concatenated  control  steers~$x^s $  to~$x^f$ at time~$T$.
In particular,  this   control
steers~$x_n(0)=x^s_n$ to~$x_n(T)=x^f_n$. Since~$u_n$ is the constant control value such that~$R^f=u_n x^f_n$,
this yields~$R(T)=
    u_n(T) x_n(T)= R^f$,
and this completes the proof.~\IEEEQED


\begin{Remark}
  Note that the construction above may lead
to a production rate $R(t)$ that is discontinuous at $t=0$.
This  can be easily overcome using any control $u_n(t)$, $t\in[0,\varepsilon]$,   that smoothly
interpolates between the value~$\frac{R^s}{x_n^s}$ at $t=0$, and the value $u_n:=\frac{R^f}{x_n^f}$ at $t=\varepsilon$.
For example, $u_n(t)$ could be picked linear in~$t\in[0,\varepsilon]$. We can then apply the constant controls $u_0,\dots,u_{n-1}$ at $t=\varepsilon$, and continue with the argument above, while noting that now we require $\tau>\varepsilon$.
\end{Remark}

{\sl Proof of Proposition~\ref{prop:ome}.}
Consider the RFM with rates~$\lambda_0,\dots,\lambda_n$.
It was shown in~\cite[Proposition $1$]{RFM_max} that $R$ is a strictly increasing function of every~$\lambda_i$.
This means that in order  to analyze~$\Omega $
in the controlled~RFM with~$u\in\U$ it is enough to consider the
reachable set for the controls~$u(t)\equiv 0$ and~$u(t)\equiv c$.
 It has been shown in~\cite{RFM_max} that for the rates~$\lambda_0,\dots,\lambda_n$, the steady state
production rate is~$R=(\zeta_{MAX}(A(\lambda_0,\dots,  \lambda_n)))^{-2}$.
Thus for the two controls above~$R(t)$ in the controlled RFM converges to~$0$ and to~$M:=(\zeta_{MAX}(A(q_0,\dots,  q_n)))^{-2}$.  We conclude  that~$\Omega(\Theta)=[0,M]$.~\IEEEQED

{\sl Proof of Proposition~\ref{prop:asyconsingle}.}
Pick~$R^f \in \Int(\Omega )$. Our goal is to show that there exist a finite time~$T\geq 0$ and a
control~$u\in\U$ that steers~$R(t)$ to~$R^f$ in     time~$T$.
We consider two cases.

\noindent {\sl Case 1.} Suppose that~$n\notin\Theta$. Since~$R^f \in \Int(\Omega)$,
 there exists~$\varepsilon>0$ such that
$(R^f-\varepsilon) \in \Omega$ and~$(R^f+\varepsilon) \in \Omega$. Therefore, there exist~$v^-,v^+ \in [0,c]$ such that
for the control~$u^-(t)\equiv v^-$ [$u^+(t)\equiv v^+$] the production rate converges to~$R^f-\varepsilon$ [$R^f+\varepsilon$] for any~$x_0$.
Applying~$u^-$ for a sufficiently long time~$T_1$
yields~$R(T_1)<R^f$.
 Now applying~$u^+$   for a sufficiently long time~$T_2$
yields~$R(T_1+T_2)>R^f$. Since~$R(t)$ is continuous, this   implies that there exists~$T \in [T_1,T_1+T_2]$
such that $R(T)=R^f$.

\noindent {\sl Case 2.} Suppose that~$n\in\Theta$, i.e.~$R(t)=u(t)x_n(t)$. The argument used in Case~1 does not hold as is
because now
a discontinuity in~$u$ yields a discontinuity in~$R(t)$. However, it is clear that we can design  a   control~$u$
by concatenating $u(t)\equiv v^-$ for~$t \in [0,T_1]$,
  then a  function of time satisfying~$u(T_1)=v^-$ and~$u(T_1+\tau)=v^+$, with~$\tau>0$,
  and finally~$u(t)\equiv v^+$ for~$t\geq T_1+\tau$,
  and that this will steer~$R(t)$ to~$R^f$ at some final time~$T$.~\IEEEQED

{\sl Proof of Proposition~\ref{prop:sens_u}.}
It has been shown in~\cite{RFM_max} that~$\frac{\partial R}{\partial \lambda_i}$ exists and is strictly positive
for all~$i\in\{0,\dots,n\}$.
Combining this with~\eqref{eq:ep} implies that~$\frac{\partial e_k}{\partial \lambda_i}$ exists for all~$k\in\{ 1,\dots,n   \}$
and all~$i\in\{0,\dots,n\}$. Pick~$i\in\{1,\dots,n-2\}$.
Differentiating~\eqref{eq:ep}  with respect to~$\lambda_i$ yields
\begin{align} \label{eq:der_ep}
                      -\lambda_0  {e}_1' & = \lambda_1
											e_1'
											    (1- {e}_2)  - \lambda_1 {e}_1  e_2' \nonumber \\&
                      = \lambda_2  {e}_2'(1- {e}_3) -  \lambda_2  {e}_2  {e}_3'   \nonumber \\ & \vdots \nonumber \\
											&= \lambda_{i-1}  {e}_{i-1}' (1- {e}_{i})   -\lambda_{i-1}  {e}_{i-1}  {e}_{i}' \nonumber \\
											&= {e}_i(1- {e}_{i+1})  + \lambda_i {e}_i ' (1- {e}_{i+1})  - \lambda_i  {e}_i  {e}_{i+1}'   \nonumber \\
											&= \lambda_{i +1} {e}_{i+1}' (1- {e}_{i+2})-\lambda_{i +1} {e}_{i+1} {e}_{i+2}'   \nonumber \\
											& \vdots \nonumber \\
											&= \lambda_{n-1} {e}_{n-1}' (1- {e}_{n})  -\lambda_{n-1} {e}_{n-1}  {e}_{n}' \nonumber \\
                    &= \lambda_{n } e_n'  \nonumber \\
										& =R',
\end{align}
where we use the notation~$f':=\frac{\partial f}{\partial \lambda_i}$.
Since~$R'>0$, we conclude that~$e_1'<0$. Now the equation~$\lambda_1
											e_1'
											    (1- {e}_2)  - \lambda_1 {e}_1  e_2'=R'$, and the fact that~$e\in(0,1)^n$  yield~$e_2'<0$.
Continuing in this fashion yields~$e_j'<0$ for all~$j \leq i$. The last
equality in~\eqref{eq:der_ep}
yields~$\lambda_n e_n'>0$, so~$e_n'>0$. Now the equality~$ \lambda_{n-1} {e}_{n-1}' (1- {e}_{n})  -\lambda_{n-1} {e}_{n-1}  {e}_{n}'  =R'$ yields~$e_{n-1}'>0$, and continuing in this fashion yields~$e_j'>0$ for all~$j > i$. This completes the proof for the case~$i\in\{1,\dots,n-2\}$. The proof when~$i\in\{0,n-1,n\}$ is similar.~\IEEEQED


\bibliographystyle{IEEEtranS}


\begin{IEEEbiography}[{\includegraphics[width=1.25in,height=1.25in,clip,keepaspectratio]{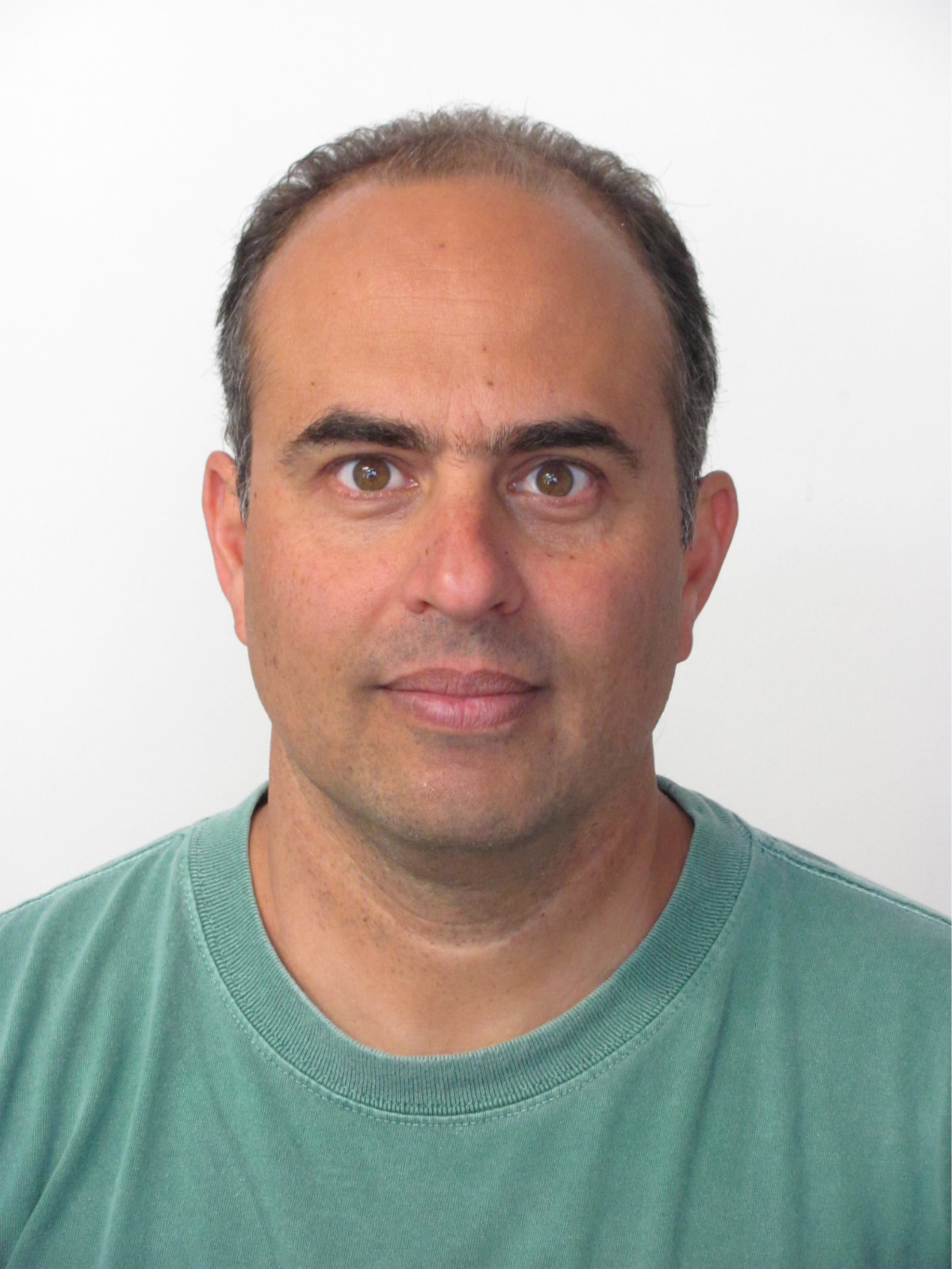}}]{Yoram Zarai}
  received the BSc (cum laude), MSc, and PhD degrees in Electrical Engineering from Tel Aviv University, in 1992, 1998, and 2017, respectively. He   is currently a postdoctoral researcher at Tel Aviv University. His research interests include modeling and analysis of biological phenomena, machine learning and signal processing.
\end{IEEEbiography}

\begin{IEEEbiography}[{\includegraphics[width=1.25in,height=1.25in,clip,keepaspectratio]{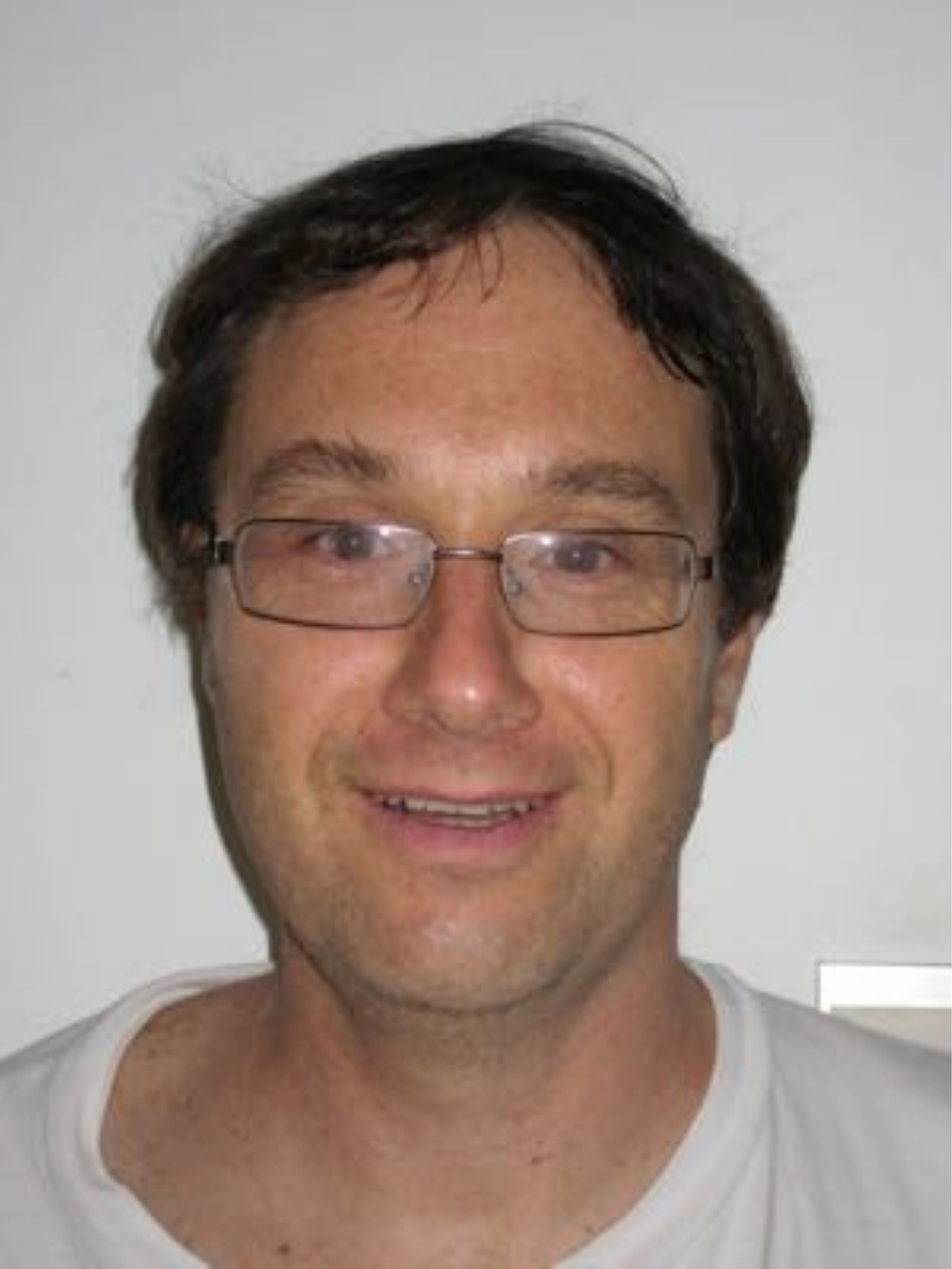}}]{Michael Margaliot}
received the BSc (cum laude) and MSc degrees in Electrical Engineering from the Technion-Israel Institute of Technology-in 1992 and 1995, respectively, and the PhD degree (summa cum laude) from Tel Aviv University in 1999. He was a post-doctoral fellow in the Dept. of Theoretical Mathematics at the Weizmann Institute of Science. In 2000, he joined the Dept. of Electrical Engineering-Systems, Tel Aviv University, where he is currently a Professor and Chair. Dr. Margaliot’s research interests include the stability analysis of differential inclusions and switched systems, optimal control theory, fuzzy control, computation with words, Boolean control networks, and systems biology. He is co-author of New Approaches to Fuzzy Modeling and Control: Design and Analysis, World Scientific, 2000 and of Knowledge-Based Neurocomputing: A 
Fuzzy Logic Approach, Springer, 2009. He currently serves as Associate Editor for the journal IEEE Transactions on Automatic Control.
\end{IEEEbiography}

\begin{IEEEbiography}[{\includegraphics[width=1.25in,height=1.25in,clip,keepaspectratio]{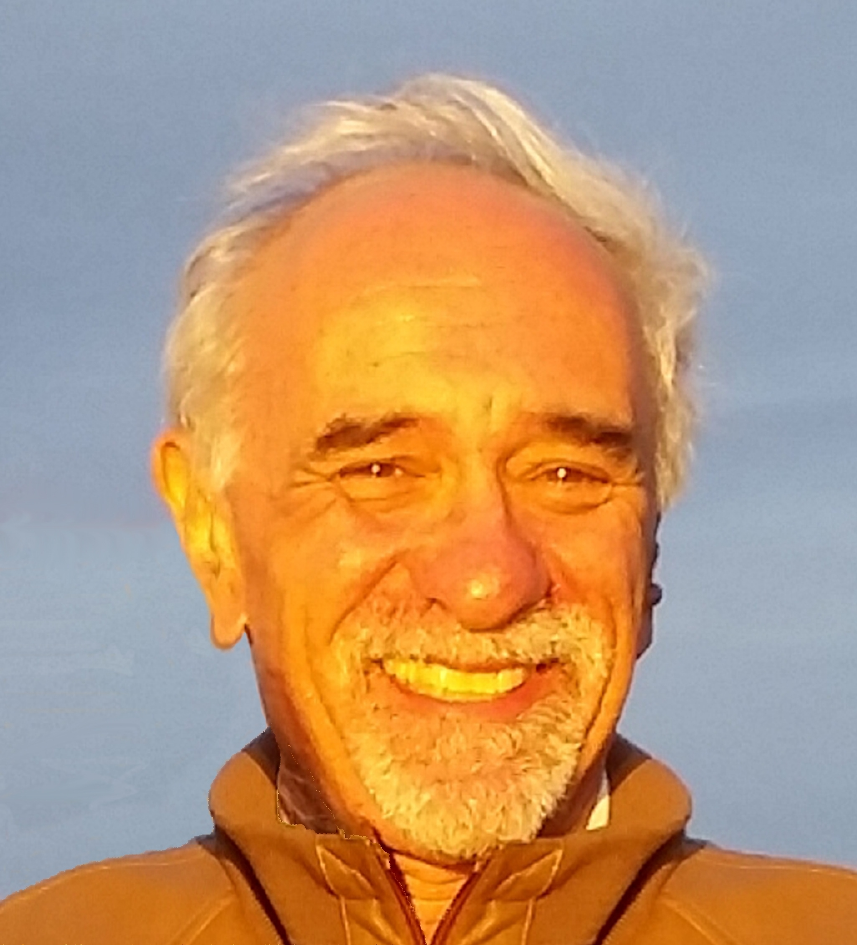}}]{Eduardo D. Sontag}
 received his undergraduate degree in Mathematics from the University of Buenos Aires in 1972, and his Ph.D. in Mathematics from the University of Florida in 1976, working under Rudolf E. Kalman.  Since 1977, he has been at Rutgers University, where he is a Distinguished Professor in the Department of Mathematics.  He is also a Member of the Rutgers Cancer Institute of New Jersey as well as of the graduate faculties of the Departments of Computer Science and of Electrical and Computer Engineering.  Sontag currently serves as head of the Undergraduate Biomathematics Interdisciplinary Major, Director of the Graduate Program in Quantitative Biomedicine, and Director of the Center for Quantitative Biology.

His current research interests are broadly in applied mathematics, and specifically in systems biology, dynamical systems, and feedback control theory.  In the 1980s and 1990s, Sontag introduced new tools for analyzing the effect of external inputs on the stability of nonlinear systems (``input to state stability'') and for feedback design (``control-Lyapunov functions''), both of which have been widely adopted as paradigms in engineering research and education.  He also developed the early theory of hybrid (discrete/continuous) control, and worked on learning theory applied to neural processing systems as well as in foundations of analog computing.  Starting around 1999, his work has turned in large part to developing basic theoretical aspects of biological signal transduction pathways and gene networks, as well as collaborations with a range of experimental and computational biological labs dealing with cell cycle modeling, development, cancer progression, infectious diseas
 es, physiology, synthetic biology, and other topics.  He has published about 500 papers in fields ranging from control theory and theoretical computer science to cell biology, with over 33,000 citations and a (google scholar) h-index of 81.

Sontag is a Fellow of the IEEE, AMS, SIAM, and IFAC.  He was awarded the Reid Prize by SIAM in 2001, the Bode Prize in 2002 and the Control Systems Field Award in 2011 from IEEE, and the 2002 Board of Trustees Award for Excellence in Research and the 2005 Teacher/Scholar Award from Rutgers.
\end{IEEEbiography}

\begin{IEEEbiography}[{\includegraphics[width=1.25in,height=1.3in,clip,keepaspectratio]{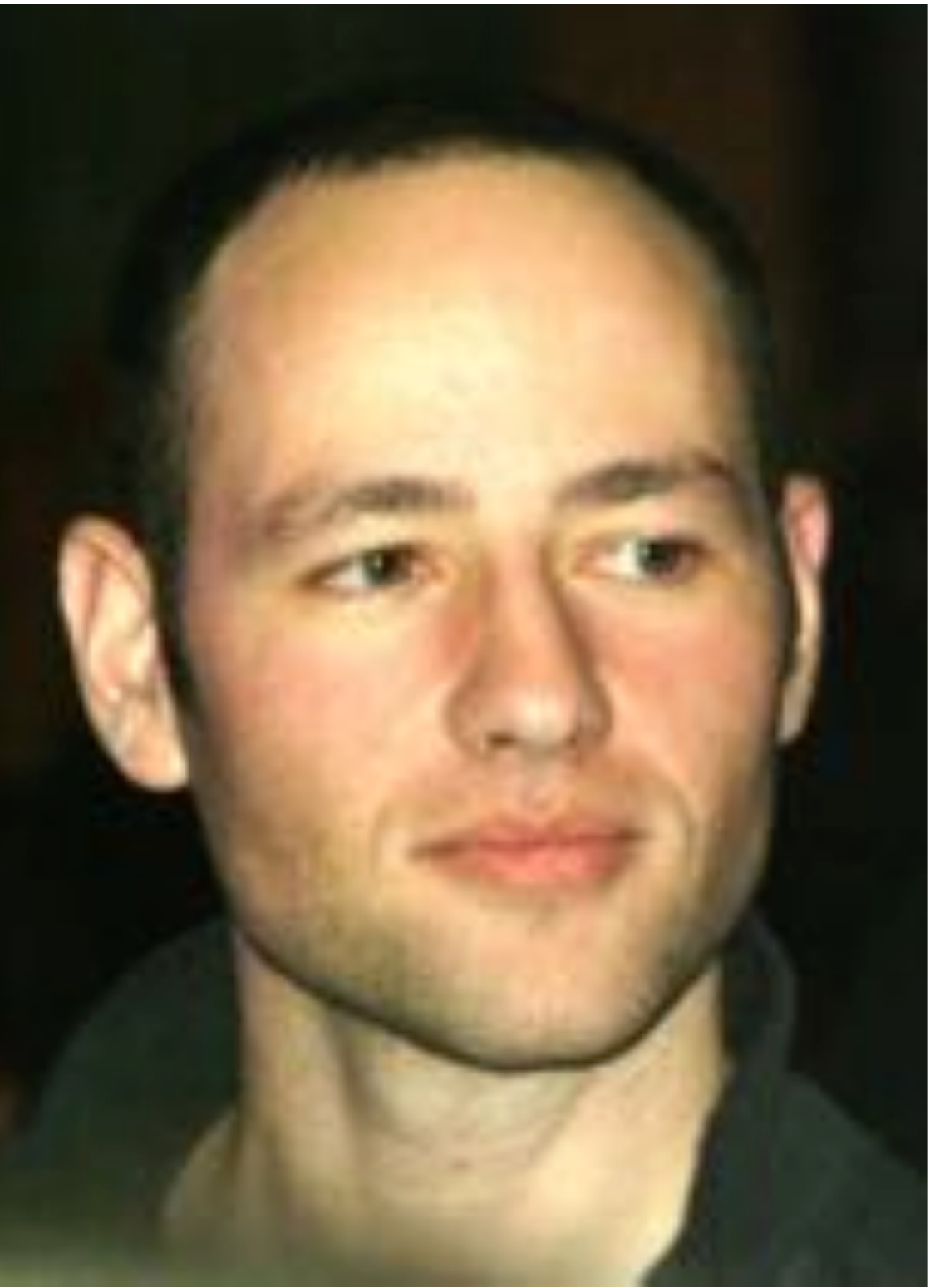}}]{Tamir Tuller}
received the B.Sc. degree in electrical engineering, mechanical engineering and computer science from Tel Aviv University, Tel Aviv, Israel, the M.Sc. degree in electrical engineering from the Technion- Israel Institute of Technology, Haifa, Israel, and two Ph.D. degrees, one in computer science and one in medical science, from Tel Aviv University. He was a Safra Postdoctoral Fellow in the School of Computer Science and the Department of Molecular Microbiology and Biotechnology at Tel Aviv University, and a Koshland Postdoctoral Fellow in the Faculty of Mathematics and Computer Science in the Department of Molecular Genetics at the Weizmann Institute of Science, Rehovot, Israel. In 2011, he joined the Department of Biomedical Engineering, Tel Aviv University, where he is currently an  Associate Professor. His research interests fall in the general areas of synthetic biology, systems biology, and computational biology. In particular, he works on deciphering, modeling, and engi
 neering of gene
 expression.\end{IEEEbiography}

\end{document}